\documentclass[acmsmall,screen]{acmart}

\AtBeginDocument{%
  \providecommand\BibTeX{{%
    \normalfont B\kern-0.5em{\scshape i\kern-0.25em b}\kern-0.8em\TeX}}}

\setcopyright{acmcopyright}
\copyrightyear{2022}
\acmYear{2022}
\acmDOI{10.1145/1122445.1122456}

\acmConference[CSCW '22]{CSCW '22: ACM Conference on Computer-Supported Cooperative Work and Social Computing}{November 12--16, 2022}{Taipei, Taiwan}
\acmBooktitle{CSCW '22: ACM Conference on Computer-Supported Cooperative Work and Social Computing, November 12--16, 2022, Taipei, Taiwan}
\acmPrice{15.00}
\acmISBN{978-1-4503-XXXX-X/18/06}

\acmSubmissionID{123-A56-BU3}

\usepackage{csquotes}
\usepackage{hyperref}
\newcommand{\sas}{TechnologyOrganization}
\newcommand{\fs}{OutreachPlatform}
\newcommand\add[1]{{#1}}
\newcommand\minoradd[1]{{#1}}
\definecolor{lightgray}{gray}{0.9}

\begin{document}

\title[Ethics and Efficacy of Anti-Trafficking SMS Outreach]{Ethics and Efficacy of Unsolicited Anti-Trafficking SMS Outreach}

%\begin{comment}
\author{Rasika Bhalerao}
\affiliation{%
  \institution{New York University}
  \city{Brooklyn}
  \state{NY}
  \country{USA}}
\email{rasikabh@nyu.edu}

\author{Nora McDonald}
\affiliation{%
  \institution{University of Maryland}
  \city{Baltimore}
  \state{MD}
  \country{USA}}
\email{noramcdonald@gmail.com}
\author{Hanna Barakat}
\affiliation{%
  \institution{Brown University and Max Planck Institute for Software Systems}
  \city{Providence}
  \state{RI}
  \country{USA}}
\email{hanna_barakat@brown.edu}
\author{Vaughn Hamilton}
\affiliation{%
  \institution{Max Planck Institute for Software Systems}
  \city{Saarbrücken}
  \country{Germany}}
\email{vaughn.j.hamilton@gmail.com }

\author{Damon McCoy}
\authornote{Both authors advised this study.}
\email{mccoy@nyu.edu}
\affiliation{%
  \institution{New York University}
  \city{Brooklyn}
  \state{NY}
  \country{USA}
}

\author{Elissa M. Redmiles}
\authornotemark[1]

\email{eredmiles@gmail.com}
\affiliation{%
  \institution{Max Planck Institute for Software Systems}
  \city{Saarbrücken}
  \country{Germany}
  }

\renewcommand{\shortauthors}{Bhalerao, et al.}
%\end{comment}




\settopmatter{printacmref=false}
\setcopyright{none}
\renewcommand\footnotetextcopyrightpermission[1]{}
\pagestyle{plain}

\begin{abstract}
The sex \add{industry} exists on a continuum based on the degree of work autonomy present in one's labor conditions: a high degree of autonomy exists on one side of the continuum where certain independent sex workers have a great deal of agency, while much less autonomy exists on the other side, where sex is traded under conditions of human trafficking.
Various organizations across North America perform outreach to sex workers to offer assistance in the form of services (e.g., healthcare, financial assistance, housing) as well as prayer and intervention.
Increasingly, technology is used to look for trafficking victims and/or facilitate the provision of assistance or services, for example through scraping and parsing sex \add{industry} workers' advertisements into a database of contact information that can be used by outreach organizations.
However, little is known about the efficacy of anti-trafficking outreach technology, nor the potential risks of using such technology to identify and contact the highly stigmatized and marginalized population of those working in the sex \add{industry}.

In this work, we investigate the use, context, benefits, and harms of an anti-trafficking technology platform via qualitative interviews with multiple stakeholders: the technology developers (n=6), organizations that use the technology (n=17), and sex industry workers who have been contacted or wish to be contacted (n=24).
Our findings illustrate misalignment between developers, users of the platform, and sex \add{industry} workers they are attempting to assist.
In their current state, anti-trafficking outreach tools such as the one we investigate are ineffective and, at best, serve as a mechanism for spam and, at worst, scale and exacerbate harm against the population they aim to serve.
We conclude with a discussion of best practices -- and the feasibility of their implementation -- for technology-facilitated outreach efforts to minimize risk or harm to sex \add{industry} workers while efficiently providing needed services. 
\end{abstract}
\maketitle
\section{Introduction}
Sex work is defined as the exchange of erotic labor or sexual services for money \cite{carol_leigh,unaids}.
\add{Sex workers} may perform various types of work including escorting, paid BDSM, sugaring, massage work, camming, stripping/erotic dancing, creating online content, phone sex work, free-styling, and working outdoors.
While each situation is different and cannot fit under a simple definition \cite{labelling,sexy_lies}, \add{work in} the sex \add{industry} exists on a continuum based on the degree of autonomy present in the worker's labor conditions: a high degree of autonomy exists on one side of the continuum where certain independent sex workers have a great deal of agency, while much less autonomy exists on the other side, where sex is traded under conditions of human trafficking, in this instance, typically termed sex trafficking \cite{continuum}.\footnote{\add{Throughout this paper, we use the term ``sex industry workers'' to refer to the population across the autonomy continuum described above.}}

Sex trafficking receives a disproportionate amount of media and legislative attention relative to other forms of human trafficking~\cite{halperin2017war}.
As a result of this focus, there has been a growth in the number of Non-Governmental Organizations (NGOs) focused on anti-sex trafficking efforts. Hundreds of technological interventions have also been designed to combat sex trafficking, often in collaboration with law enforcement and/or anti-sex trafficking NGOs \cite{305_services,dimas2021survey}.
For example, DARPA Memex\footnote{\url{https://www.darpa.mil/about-us/timeline/memex}} is a program funded by the United States Department of Defense to develop technologies that index ``forums, chats, advertisements, job postings, hidden services'' related to the sex \add{industry} in an effort to identify human trafficking online.
Similarly, Spotlight\footnote{\url{https://www.thorn.org/spotlight/}} is another technology application developed by Thorn, an NGO that aims to stop \add{the} sex trafficking of minors by scraping online sex \add{industry}-related advertisements and forwarding them to law enforcement \cite{thorn}.
Finally, Freedom Signal, a platform built by NGO Seattle Against Slavery,\footnote{\url{https://www.seattleagainstslavery.org/}} connects providers of direct services with sex \add{industry} workers who advertise their services online to ``build trust with vulnerable populations in acute crisis.''\footnote{\url{https://freedomsignal.org/}}

While such technologies aim to identify trafficking victims and provide needed services such as medical assistance, financial assistance, housing, immigration assistance, and mental health resources, little data exists on how effective such anti-trafficking technologies have been in addressing sex trafficking. 
%One of the few studies of an anti-trafficking intervention program, New York's Human Trafficking Intervention Initiative, found that while the program provided services to over 2,400 people who would have otherwise been prosecuted for sex-work-related charges (e.g., loitering, prostitution), only seven people have been charged with sex trafficking during the same duration. The program is effective at identifying sex workers but not necessarily sex traffickers~\cite{halperin2017war}. 
\add{Most forms of sex industry work are criminalized in most of North America.\footnote{The exception is some counties in Nevada in the United States, where one specific type of sex work is legalized and heavily regulated. There are also some state decriminalization bills under consideration at the time of this writing. Camming, pornography, and stripping are currently regulated.}
In addition to harms specific to the criminalization of sex \add{industry} work (Section~\ref{position_statement}), holding an illegal job in North America leaves one subject to fines, without sufficient health insurance, and vulnerable to police brutality \cite{hoefinger,sexhum,decrim,walking_while_trans_repeal}.}
Sex work is highly stigmatized, and those involved in the sex industry are disproportionately targeted by law enforcement \cite{carceral_fosta} even in countries where sex work is legal and despite the decriminalization of sex work being supported by the World Health Organization, the United Nations, and other international human rights bodies~\cite{halperin2017war}.
Enforcement also disproportionately targets marginalized individuals based on their gender or sexual identity and race \cite{responding,andrijasevic}.
Further, sex work is also often equated with sex trafficking under the premise that all labor in the sex \add{industry} is coercive~\cite{halperin2017war,2020_annual_report}.
Thus, anti-sex trafficking efforts, including technology-facilitated interventions that identify and catalog sex \add{industry} workers \add{regardless of their working conditions or autonomy}, can pose significant risks to sex \add{industry} workers' digital and physical safety, which is already precarious~\cite{elissa1, elissa2,FOSTA_musto}.
% Numerous studies in the last few years have demonstrated that technology reflects, reproduces, and exacerbates structural inequality \cite{algorithms_of_oppression,dirty_data}.

In this study, we focus on the use, context, benefits, and harms of one specific anti-trafficking technology: \fs, developed and maintained by \sas.
(\fs\ and \sas\ are pseudonyms used to maintain anonymity. The original names indicate a strong anti-trafficking focus.)
We sought to understand whether and how outreach technology in the form of mass unsolicited SMS messages can be used to assist victims of trafficking while mitigating or reducing harm caused to other recipients of outreach.
To do so, we took a multi-stakeholder view, conducting interviews with the developers of \fs\ (n=6), organizations that perform outreach to sex \add{industry} workers (n=17), and \add{sex industry workers} who had received outreach from anti-trafficking NGOs or wished to receive outreach (n=24).
The members of the sex industry that we interviewed ranged from sex workers with a high degree of agency and autonomy over their working conditions to one survivor of severe sex trafficking, with many in between.
We specifically recruited sex \add{industry} workers who have been contacted by anti-trafficking organizations and some who have not but would like to be contacted. 

We report on the misalignment between the goals of the outreach technology, organizations using the technology, and sex \add{industry} workers---e.g., organizations' misunderstanding of the ideal beneficiaries of unsolicited outreach messages and thus their language, timing, and interpretation of SMS messaging interactions.
We find that \add{sex industry workers across the continuum} need resources that some organizations can provide, but changes to the technology platform and outreach practices are required to deliver them in a way that transforms them from harmful spam into messaging that is only a minor nuisance to those who are not intended recipients.
We share our findings on the potential harms of technology-facilitated outreach and discuss the implementation of mitigation strategies.

The CSCW community is increasingly acknowledging the need for research to explore these types of misalignments \cite{nora_open_collaboration} and include marginalized individuals in the design process \cite{constanza_chock}.
This research aims to bridge this gap and contribute to frameworks for future CSCW and CHI research.
While technologists sometimes believe that technology is ``unbiased,'' we demonstrate that it reflects the values of its creators and the structural biases and discriminatory practices inherent in how data is used \cite{feminist_hci}.
\add{Given the paradigm of a narrow focus on} sex trafficking, data that is scraped from websites reflects the assumptions of developers (about what these sites represent) and the structural inequalities of the criminalized landscape from which the data is harvested (not unlike \citet{dirty_data}).
The assumptions and the data that are the basis for sex trafficking outreach tools, like the one we studied, are shaped by a system that pushes sex work underground and classifies all sex \add{industry} workers as ``[potential] victims'' by default.
We find that even organizations who express sensitivity to the difficulty of distinguishing non-trafficking situations from human trafficking may be committed to their mission to ``save'' \add{sex industry workers}---even if that means putting at risk or inflicting harm on the majority of their outreach recipients.

\section{Background}
\subsection{Defining Sex Work, Sex Trafficking, and Anti-Trafficking Organizations}

Common definitions of human trafficking include migrant participation in any industry under force, fraud, or coercion \cite{Palermo_Protocol} or under debt bondage \cite{obama_law}.
Some additionally emphasize participation in the sex \add{industry} as a minor \cite{Palermo_Protocol} or forced marriage \cite{ilo_forced_marriage}, while other definitions include forced or low-paid labor in the prison industrial complex \cite{polaris_prison,prison_exploitation}.
Sex trafficking is a type of human trafficking that involves sexual exploitation \cite{labor_trafficking_prevalence,sage_perspectives}.

Statistically, the population along the continuum of \add{the sex industry} contains relatively very few sex trafficking victims \cite{andrijasevic,davidson,kempadoo1,kempadoo2,revolting_prostitutes,playing_the_whore,kenway,brokered_subjects,responding,shifting_perspectives,decrim,labor_trafficking_prevalence,sage_perspectives}.
One organization that studies sex work and trafficking in Canada, and provides support to \add{sex industry workers}, suggests that the distribution of people on the sex \add{industry} continuum fits a bell curve: the vast majority see sex work as a ``job,'' and a very small number see it as either ``survival'' or a ``career'' \cite{shifting_perspectives}.
An individual's place on this continuum primarily depends on their ``options, choices, resources, and privileges'' \cite{shifting_perspectives}. 
% Some sex worker scholars feel there is no autonomy under capitalism due to limited ``freedom'' \cite{marxism}; we choose to use the spectrum of autonomy for our definitions because it is commonly understood in sex work literature \cite{sage_perspectives} and a convenient way of conceptualizing human behavior.

In this study we focus on technology used by organizations whose purpose is to assist victims of sex trafficking and/or provide resources to \add{sex industry workers} in the form of services such as medical assistance, financial assistance, housing, immigration assistance, and legal assistance.
These organizations perform outreach to \add{sex industry workers} to provide concrete services needed by people all along the continuum.
For this study, we categorize such organizations into three groups: rescue organizations, sex-worker-led organizations, and social service organizations.
``The rescue industry'' (term coined by \citet{sex_at_the_margins}) describes the constellation of government agencies, NGOs, and businesses that serve them.
Rescue organizations generally support \add{prohibition of} the sex industry, and they tend to be well-funded, second-wave feminist or religious groups \cite{sexy_lies,sexhum,sex_at_the_margins,gaatw_agustin_interview,sage_perspectives}.
Sex-worker-led organizations\footnote{E.g., The Sex Workers Outreach Project (\url{https://swopusa.org}).} overwhelmingly support decriminalization of sex work and propose that increased worker rights would reduce exploitation in the sex industry.
Social service organizations are not specific to the sex \add{industry} and include organizations such as sex therapy clinics, crisis hotlines, women’s rights organizations, and government jurisdictional organizations.

Prior research finds concrete differences between rescue organizations and sex-worker-led organizations through content analysis of their websites: the rhetoric on rescue organization websites tends to focus on fighting trafficking and helping people \add{``exit''} the sex \add{industry} (with or without religious undertones), and the rhetoric on sex-worker-run organization websites tends to focus on harm reduction or the sex \add{industry} worker community, while offering similar resources \cite{sexy_lies}.
Prior work shows that many sex \add{industry} workers feel misrepresented by rescue organizations, who leverage a rare, extreme example of trafficking to characterize the entire sex industry \cite{natasha_gordon,natasha_zhang,natasha_davies}.
As a result, sex \add{industry} workers have called for ``a more nuanced appreciation of the relationship between agency and victimization'' in anti-trafficking literature and organizations \cite{labelling}.
Sex \add{industry} worker activists and journalists have investigated the efforts of some rescue organizations, finding that publicized news articles describing past successes in rescuing sex \add{industry} workers covered up misuse of donations and lack of real help to sex \add{industry} workers \cite{cupcake_girls}.
Some rescue organizations work closely with law enforcement to forcibly remove workers from the sex industry \cite{sex_at_the_margins}.
Sex-worker-led organizations have shown that organizations that consider sex \add{industry} workers as ``potential victims'' and encourage them to ``abstain from sex work'' are unable to help most victims of trafficking but instead end up causing harm to sex \add{industry} workers through stigma rather than reducing harm \cite{harm_reduction}.
Other evidence of many sex \add{industry} workers' perception of the rescue industry can be found in memes (Appendix~\ref{appendix:meme}) and Ted Talks \cite{juno_ted}.

\subsection{Outreach Technology}

\begin{figure}%[15]{L}{0.6\textwidth}
    \includegraphics[width=0.6\textwidth]{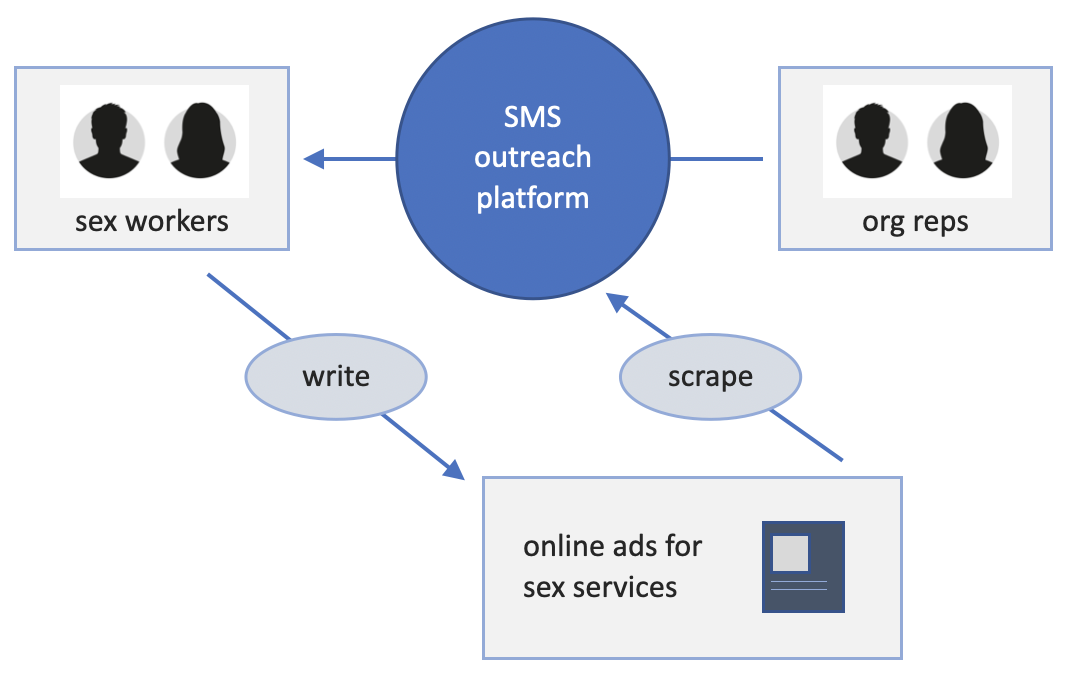}
    \caption{Overview of the \fs\ technology.}
    \label{fig:tech_overview}
\end{figure}

In this study, we focus on the technology that organizations use to perform outreach to \add{sex industry workers}.
Organizations do outreach in many ways, including:
\begin{itemize}
    \item operating 24-hour hotlines which anyone, including sex \add{industry} workers, can call;
    \item receiving referrals from places such as medical service providers, legal courts, youth-centered groups, and other organizations similar to themselves;
    \item proactively going to strip clubs, massage parlors, and other places where sex \add{industry} workers are employed;
    \item proactively contacting sex \add{industry} workers through ads that \add{are posted online by sex industry workers to attract clients}.\footnote{The word ``client'' has different meanings based on context in our study. \sas's clients are the organizations, who also provide the majority of \sas's funding. The clients of organizations are sex \add{industry} workers and survivors of sex trafficking, and those organizations' funding comes from a variety of sources including grants and donations. The clients of sex \add{industry} workers are people who purchase erotic labor.}
\end{itemize}
In particular, we looked at organizations' use of the service provided by \sas: \fs.\footnote{\sas\ is a 501(c)(3) non-profit organization that has created and operated \fs.}
As shown in Figure~\ref{fig:tech_overview}, \fs\ scrapes advertisements for sex work that are posted on dedicated sex industry marketing websites, collects the phone numbers and demographic details (gender, age, location, and the timestamp at which the ad was last updated) from the text of the advertisements, and then imports this information into the \fs\ database.
Organization representatives can then use \fs's functionality to communicate via SMS with the people whose phone numbers are in the database, starting with a mass SMS text message to selected phone numbers.
Then, on a second screen, organization representatives can continue and manage conversations with anyone who replied to the initial mass text outreach.
\fs\ offers additional functionality to organization representatives to filter contact information by demographic details; see Section~\ref{identity} for findings on how this filtering is used in practice.

% Before Freedom Signal was developed, \sas\ acquired and ran Project Intercept.
% Project Intercept only scraped women's sex ads and also had an additional component that Freedom Signal does not: a demand reduction bot that posted fake advertisements to sex ad sites and helped law enforcement officers to catch people who tried to purchase trafficked sex.
% According to \sas\ employees we interviewed, the demand reduction program was ended for a multitude of reasons including that those caught buying sex from the bots did not match the racial demographics of the intended targets, \sas\ did not want to contribute to the cycle of incarceration, and the product did not have a good market fit as compared to a competitor.

\section{Related Work}
Prior studies of technology use in the sex industry explored sex workers' use of digital platforms~\cite{internet_sex_work,jones2020camming} and payment mechanisms~\cite{sanders2020paying}, security considerations that sex workers take into account to do their jobs~\cite{elissa2}, and the digital discrimination they face~\cite{elissa1,hacking_hustling_erased}.
Additional prior work has explored the harms of technology-related anti-trafficking legislation such as FOSTA/SESTA \cite{FOSTA_musto,carceral_fosta,hacking_hustling_erased}, the stigma of sex work online\footnote{\url{https://medium.com/berkman-klein-center/what-can-tech-learn-from-sex-workers-8e0100f0b4b9}} \cite{garcia_thesis}, and the technology used by sex workers' rights organizations for social justice-related services \cite{fighting_stigma_saving_lives}.
These studies \add{and news articles explain how} criminalization and stigma put sex \add{industry} workers in precarious and often vulnerable positions, and made suggestions on ways to design technology to mitigate the harms unique to the sex industry; the general consensus was that including sex \add{industry} worker input in the design process of technological tools would help accommodate diversity, privacy, safety, and ease of use \add{\cite{sesta_kill,bills_cracking,backpage_crucial,repeal_sesta,pushed_off}}.
While this prior work has studied how sex workers use technology and how technology regulation impacts sex workers, no prior work has examined how technology platforms designed to scale provision of assistance to sex \add{industry} workers and anti-trafficking efforts impacts the stakeholders involved in and affected by such technological systems.
Given the growing interest in developing such systems~\cite{doj_pr_2020}, studying these technologies through an HCI lens is critical to informing the ethical and effective development of such systems.

\subsection{Anti-trafficking Technology}

Prior work categorized and quantitatively analyzed hundreds of technological and analytical anti-trafficking tools and found that these tools are failing to address trafficking~\cite{305_services} and that 
``new methods will need to be developed to account for the various social nuances inherent to HT [human trafficking]'' \cite{dimas2021survey}.

Our qualitative work aims to address these social nuances relevant to anti-trafficking technologies by understanding the goals and experiences of all stakeholders (technology developers, users of anti-trafficking technologies) and the community reached by the technology (sex industry workers).
Most related to our work, \citet{chen_usenix_rescue} investigated the technological security risks of ``victim service providers'' (VSPs) -- social workers and volunteers working at anti-trafficking organizations. Of the 11 organizations interviewed, only one conducted unsolicited outreach of the type we examine in this work; as such, their findings primarily focus on post-outreach technological interactions between service providers and clients. 
In their findings on these interactions, \citet{chen_usenix_rescue} found that VSPs are balancing ``building trust with their clients (often by giving them as much autonomy as possible) while attempting to guide their use of technology to mitigate risks around revictimization'' \cite{chen_usenix_rescue}; not unique to this study is the problematic framing of the return to sex work as ``revictimization'' and the lack of full support for sex \add{industry} worker autonomy.

We expand on this prior work in two directions. First, we are, to the best of our knowledge, the first study to focus on investigating anti-trafficking outreach technology used for unsolicited outreach to sex industry workers. Second, we examine the goals and impacts of anti-trafficking outreach technology with regards to all stakeholders involved in the outreach ecosystem. We believe this second contribution is critical to interpreting the ethics and efficacy of a significant part of the anti-trafficking technology ecosystem: anti-trafficking outreach technologies.

\add{\subsection{Unsolicited Outreach Technologies}}

Outside of the trafficking context, \citet{hiv_outreach} investigated the barriers preventing sex workers from taking HIV tests in the European Union (E.U.), including a study of a variety of ways that healthcare providers or social workers \add{used technology to perform} outreach.
Even in the E.U. where sex work is legal and regulated (albeit in a variety of ways by different countries), they found barriers to sex worker \add{response to outreach messages}, including \add{recipients' fear that law enforcement is involved in the communication, not knowing whether the sender is friendly towards sex workers,} lack of accessibility, past negative experiences, and lack of health education \cite{hiv_outreach}.
A pilot healthcare outreach method that involved manually sending SMS messages found that the recipients of outreach text messages often did not trust the outreach, and that health workers sending the messages were worried about reaching ``third-party operators'' preventing them from reaching their intended recipients \cite{hiv_outreach}.
While \fs\ can be seen as an answer to their call for ``scaling up'' the amount of technology-facilitated outreach to sex workers online and of the diversity of its recipients \cite{hiv_outreach}, we \add{set out to find if} it also scales up the issues of unsolicited SMS outreach.

Similarly, \citet{medical_scoping_review} reviewed a variety of studies of ways that sex workers use ``electronic occupational health and safety tools'' or that organizations external to the sex industry use these tools to contact sex industry workers.
They measured the efficacy of SMS and online chat outreach encouraging occupational health practices (HIV testing, condom use, etc.) from non-sex worker-led organizations to sex industry workers in several countries including the United States, China, India, Mexico, the United Kingdom, Canada, South Africa, Zimbabwe, and Mozambique.
They found that stigma, the risk of family members inadvertently seeing a notification, and laws criminalizing sex work prevented such outreach from being fully effective, and that non-sex worker-led organizations benefited from the input of sex workers in designing outreach \add{methods and messages}.
They ``recommended that future research involve sex workers in the formulation of the research question and scope of the study to accurately identify the issues that sex workers face in their personal and professional lives\add{,'' supporting the idea that technology designed for marginalized populations should be designed with input from, if not led by, members of that population~\cite{design_justice}.}

% \add{Unsolicited communications are also used for other purposes such as activism in repressive states.
% \citet{arab_spring} discusses how social media was a powerful tool in the ``Arab Spring'' to disseminate information and recruit ``insurgents,'' leading to physical action later on if it was successful.
% They emphasize the importance of ``clandestine communication'' to advertise to a vast audience in order to reach the few who will join their cause at great personal risk.
% }

\add{
Mass unsolicited messaging is not limited to NGO outreach. Most prior work on mass unsolicited messaging is focused on marketing campaigns.
\citet{permission} found that recipients' ``willingness to give permission to receive SMS advertisements''  was not correlated with brand familiarity; rather, ``the possibility to withdraw at any time'' and ``personal data disclosure only with consent'' were key.
\citet{reconciliation} concluded that marketing campaigns should use various messaging platforms, not just SMS, to personalize communication and foster a sense of trustworthiness.
\citet{attitude} recommended the discontinuation of their unsolicited SMS campaign because the messages lacked informational value and relevance since ``the messages were not permitted by the subscribers that receive them, which is why they delete them immediately after receiving them.''
\citet{aida} found ``a general negative response to unsolicited SMS adverts as majority of the sample reported no attention, no interest, no desire and no action to unsolicited SMS adverts.''
}

%% BELOW ARE ``MEANER'' EXAMPLES WE CUT %%

% The design of anti-trafficking tools will benefit greatly from the inputs of sex workers and survivors of trafficking.

% Insights gained from the input of survivors of commercial sexual exploitation of children (CSEC) include ``youth often do not self-identify themselves as victims,'' ``arresting victims undermines efforts to combat CSEC,'' and ``civil liberties are important considerations'' \cite{minors_berkman}.

% Studies such as \citet{wivi_iot} should raise alarms for those who are concerned about unintended use, consequences, and risk to other vulnerable populations: to find people being trafficked, they use technology that can ``see through walls'' to catch people concealed in rooms, sealed containers, and boxes.

% We urge projects such as the ``Hacking for Freedom: A Hackathon to Stop Sex Trafficking'' hackathon\footnote{\url{https://news.mit.edu/2018/mit-sloan-executive-mba-grads-hacking-freedom-to-fight-human-trafficking-1204}} to include sex industry representatives for guidance and context on law enforcement and sex trafficking \cite{garcia_thesis}.

% \href{https://scholar.google.com/scholar?hl=en&as_sdt=0%2C48&q=nlp+human+trafficking&btnG=}{Any quick search}
% for anti-trafficking technology studies will reveal a plethora of studies lacking context about trafficking, unaware of the historical conflation of sex work and trafficking, and citing problematic reports that conflate the two.

% Microsoft created a chatbot that posted fake online sex advertisements and scolded those who responded \cite{microsoft}; a prior literature review would have suggested ways that this could harm sex workers.

\add{\subsection{Designing Technology for Marginalized Populations}

Designing technologies with marginalized populations in mind is a priority in CSCW literature.
As explained by \citet{design_justice}, ``far too often, user personas are created out of thin air by members of the design team (if not autogenerated by services like Userforge), based on their own assumptions or stereotypes about groups of people who might occupy a very different location in the matrix of domination.''
Technology for social good is almost by definition biased by designers' relationship to power, which is different and likely privileged.
\citet{design_justice} argues that designers must engage with the experiences of marginalized individuals who are less visible or invisible under heteropatriarchy and capitalism, settler colonialism, racism, sexism, ableism, and other forms of structural inequality.
As an example of this problem,
\citet{design_justice} critiques disability simulation methods  arguing that the oppressed user must be engaged with or best, centered. \citet{promise_of_empathy} argue that when researchers indulge these types of empathy thinking models,  they are predisposed to respond to their own experiences and subvert those of the other. That is, if something seems real, we may be even more inclined to distance ourselves from the experience, to ``turn it off.''

Extensive ethnographic study by \citet{OLPC} of the One Laptop Per Child (OLPC) project provides a case study of how technology design for social good's best intentions are thwarted by this kind of thinking. Ames documents how laptops were introduced with the hope of changing the lives of children across the Global South, but were designed in the image of U.S. developers and not a heterogeneous group of young children with different cultures, contexts, and orientations to technology.
\citet{surreptitious_communication_design} is similarly critical of the ``design for good'' (or D4G) movement as being commercial and far removed from the realities of the individuals it is seeking to help and the challenges of social change that come with such efforts.
\citet{surreptitious_communication_design} is particularly instructive to our work, proposing the ``surreptitious communication design'' (SCD) framework for designing communication systems and campaigns in threatening contexts, where messages can be understood by vulnerable, stigmatized groups but not by those that might mean them harm.
SCD employs methods of obfuscation and ephemerality, borrowing from cryptography (e.g., coding and cloaking).
Examples of cloaking include placing messages in ways that they are unlikely to be intercepted by adversaries, such as alternative media or a format that can only be seen by people of a certain height (in child domestic abuse campaigns).}

\vspace{10pt}
Our work builds on the existing academic literature \add{on anti-trafficking technologies, outreach to sex industry workers, and unsolicited outreach.}
We take a technology-centric qualitative approach of inquiry to understand anti-sex-trafficking technologies through the lens of each of the  stakeholders of the technology.
\section{Study Design}
\label{method}
In this work, we seek to understand whether and how anti-trafficking technology via unsolicited messages can assist sex industry workers across the spectrum of autonomy.
To answer this question, we collect empirical evidence from multiple stakeholders of existing unsolicited messaging anti-trafficking technologies: the developers of those technologies, the organizations that use those technologies, and the sex \add{industry} workers who may be contacted through these technologies. We focus specifically on organizations' proactive technologically mediated outreach methods, such as using \fs\. 
This platform is recently developed, scalable, and reproducible.

In this section, we describe our interviews with each set of stakeholders: the creators of \fs\ (n=6); social workers and volunteers who do outreach to sex \add{industry} workers (n=17); and sex \add{industry} workers (n=24).
% We also describe our follow-up interviews with \sas\ and a small sample (n=3) of our original organizational interviewees, in which we present our initial findings.

\subsection{\fs\ Developers}

\subsubsection{Recruitment}
\sas\ employees and former interns who assisted with the development of \fs\ were recruited for interviews via email.
Our recruitment email requested participants for an interview regarding \fs\ and about \sas\ itself.\footnote{Example: ``We are looking to interview individuals who helped to build \fs\ to get a holistic view of the technology and the problem it's meant to solve. Would you be available to connect for a 1 hr interview?''}
All six \sas\ members that we requested agreed to be interviewed.
%Those interviewed were compensated with a \$25 Amazon gift card.

\subsubsection{Interview protocol}
We started these 1-hour long interviews by asking participants about their role in \sas\ and in developing \fs, to determine how the platform and its goals were shaped.
We then queried participants about how the design of \fs's UI and backend / data storage relates to the goals of \sas\ and the organizations that use \fs.
We also asked about \fs's goals, and how the platform makes use of identity and demographic information.
We then asked if the platform has any safety features (e.g., obfuscating either party's phone number or enforcing requests to end communication).
Finally, we asked about the organizations using the platform, and how \sas\ balances the needs and requirements of these organizations when determining \fs's goals and platform features.
%We ended the interview with an open ask for anything relevant left out.
The full interview protocol is in Appendix \ref{sas_protocol}.

\subsection{Organization Social Workers and Volunteers}

\subsubsection{Recruitment}
Most social workers and outreach volunteer participants were recruited via email, while a few were recruited via contact forms on organization websites.
Because we are interested in the technology used for anti-trafficking efforts, we focused on organizations that used \fs\ or related outreach methods.
We interviewed 17 representatives from 14 organizations; of these, 12 representatives from 11 of the organizations used \fs.
All organizations also used other technology or analog methods to facilitate their outreach.
To recruit organizations that align with \sas's values (mainly that the organizations are not purely prayer groups or law enforcement\footnote{Most organizations in our study refer sex \add{industry} workers to law enforcement if the worker reveals their age as below 18, but not otherwise.}), a point of contact at \sas\ sent out our recruitment email to organizations that do and do not use their platform, with about a 50\% response rate.
The organizations that did not respond were contacted again through their websites and advertised email addresses; at least two organizations signed up following our direct contact. 

To avoid priming, recruitment messages asked for interviews ``to learn about the work of organizations such as yours [theirs].''
To encourage participation, the message also said that all results will be anonymized and presented in aggregate and that participants could share to their comfort level.
%Participants were compensated with a \$25 Amazon gift card.

\subsubsection{Interview protocol}
We asked organization participants about their work, their organizations, and, if applicable, their relationship with \sas.
We then queried how they used technology to conduct outreach.
We probed their safety concerns for themselves and the people they reach out to, if any.
We asked about whom their organizations aim to support through outreach, how effective they are or believe they are in delivering support, and the goals they seek to accomplish through outreach.
We also asked whether and how they account for aspects of identity, including race, nationality, gender, and sexual identity.
%Finally, interviews again concluded with an open ask for any additional information they wanted to share.
The full interview protocol is in Appendix \ref{org_rep_protocol}.

\add{\subsection{Sex Industry Workers}}

\subsubsection{Recruitment}
\begin{figure}[ht]
    \centering
        \begin{minipage}{.47\textwidth}
        \centering
        \includegraphics[width=0.95\linewidth]{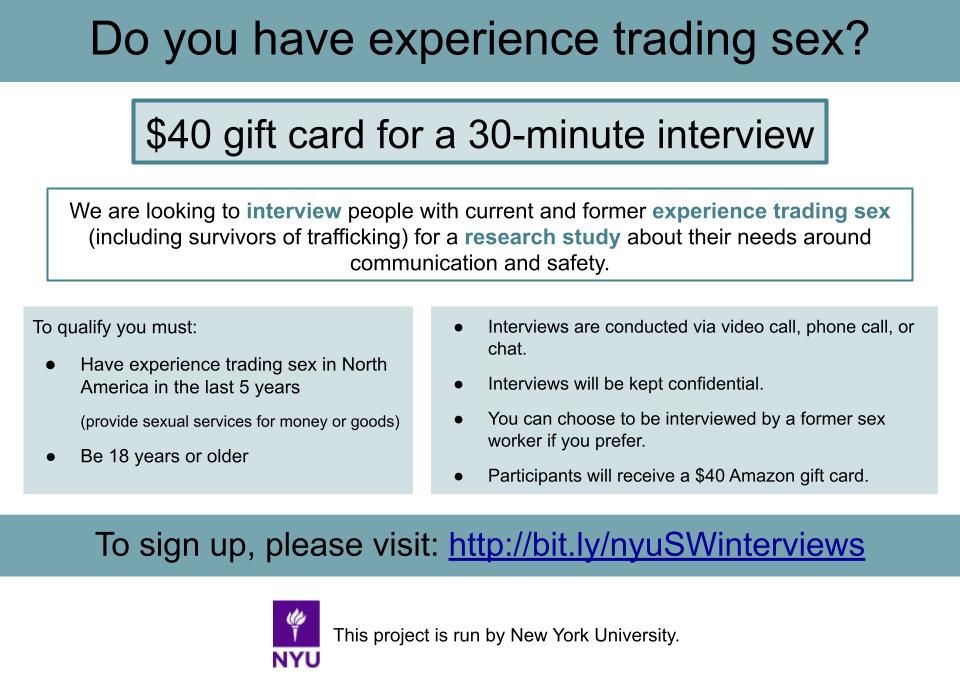}
        \caption{Flyer distributed to organizations to recruit sex industry workers in contact with them. Portions are greyed out for anonymity.}
        \label{fig:flyer_sex_trade}
    \end{minipage}%
    \hfill
    \begin{minipage}{.47\textwidth}
        \centering
        \includegraphics[width=0.95\linewidth]{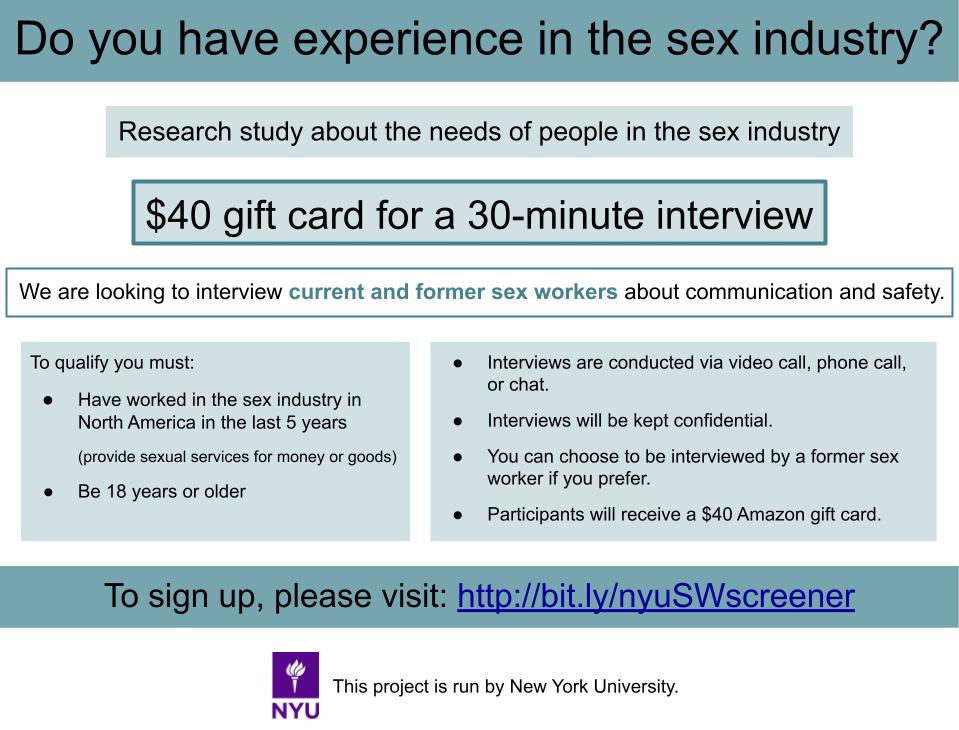}
        \caption{Flyer distributed on social media to recruit sex industry workers. Portions are greyed out for anonymity.}
        \label{fig:flyer_sex_work}
    \end{minipage}
\end{figure}

Workers in the sex industry were recruited by sending the flyer in Figure \ref{fig:flyer_sex_trade} to organizations including \sas's contacts, participants in the organizational interviews, and other rescue or sex-worker-led organizations to distribute, and by posting the flyer in Figure \ref{fig:flyer_sex_work} to social media.
\add{
The flyer posted to social media was publicly viewable (without any platform-specific targeting strategies to find sex industry workers) and tailored so as to avoid an emphasis on trafficking, since it might have otherwise been ignored by sex industry workers.
However, some survivors of trafficking who are still in contact with service organizations do not identify as sex workers, so we used the flyer that recruited ``people with current and former experience trading sex'' to match the vocabulary of this community.}

\subsubsection{Interview protocol}
We began the interviews by emphasizing that interviewees did not need to answer any question or describe any negative experience that they did not want to.
Then, we asked about the participant's background, including the type of sex work they did and why they started doing sex work.
If appropriate (i.e., for those who were not trafficked), we also inquired about how participants found, scheduled, and communicated with clients, to understand how organizations' modes of outreach aligned with workers' existing means of communication.
We then queried participants' experiences and sentiments regarding being reached out to by organizations and any recommendations they had for how the outreach could have been improved.
We then discussed participants' broader experiences seeking support or services when they needed them.
We also asked about any experiences participants had offering support to others in the sex industry.
Finally, we asked whether they felt that aspects of identity were important to the support and services they received, and if so, in what ways.
%As in the other interviews, we concluded with an open-ended ask for things they would like to share about the process of seeking support or resources, or having people reach out to them offering those resources.
The full interview protocol is in Appendix \ref{sw_protocol}.

\subsection{Data Analysis}
Developer and organization interviews were conducted via video or audio call. One organization representative requested a text-only interview (and we obliged).
Sex \add{industry} worker interviews were conducted via video or audio call, or via text chat, depending on their preference.
All three options used the WebEx conferencing system and interviewees gave consent for recording.
All interviews were recorded and/or manually transcribed.
%The text transcripts of text-chat-only interviews were also recorded.
Five of the authors conducted the interviews, took notes during and after interview sessions, and wrote memos to review each interview. Interviewers read transcripts and/or reviewed notes from all of the stakeholders' interviews. They then divided transcripts among themselves and conducted a round of open coding, producing an initial set of codebooks. 
The first author developed a final codebook based on the initial codebooks and regular discussions with the other reseachers. The five interviewers applied this final codebook to a set of transcripts (divided among interviewers). The authors regularly met to discuss, refine, and diagram the themes presented in this paper. %, choose where to focus subsequent recruitment efforts, and determine which questions from the interview protocols to focus on during interviews to achieve our study goals.
This process most closely resembles thematic analysis \cite{braun_clarke}.
% We importantly take a phenomenological stance when it comes to exploring perceptions of safety, privileging the realities and experiences of participants, regardless of whether we believe that they are, in any given situation, appropriately attuned to safety-related issues, sex-work-related definitions, or our ethical beliefs \add{Elissa says: I feel like I like this, but I don't quite understand it? Rasika says: Actually we decided to stop doing this right? @Nora help. Nora: Ha, I was just getting back to this bc yes, we can't claim we took this stance anymore. While I like to claim it with marginalized investigations in particular because it denorms the read, we very much took a stance :)}.

\subsection{Logistics, Ethics, and Anonymity}
Ethically conducting and analyzing interviews by preserving the anonymity and confidentiality of our interview participants is of utmost importance to us.
All findings, stories, and concerns noted in this study originated from more than one participant unless indicated otherwise.
Social media sites, messaging applications, and other platforms are not named in this paper to avoid any repercussions to sex \add{industry} workers using those platforms.
Each interview participant was given a consent agreement detailing the interview process as part of the confirmation for scheduling.
Scheduling of interviews \add{for all stakeholders} was facilitated by Calendly\footnote{https://calendly.com} to avoid linking personally identifiable information with interview content.
Events scheduled by Calendly included links to the authors' WebEx personal meeting room.
The WebEx meeting room could be accessed by anyone who had the link or phone number with an access code (though the authors ``locked'' it when an interview was happening) and did not require interview participants to provide personal information to join.
Interview notes, recordings, and transcripts identified participants using random anonymized IDs; in this paper we refer to sex industry workers, organization representatives, and platform developers as \add{SW1-24, ORG1-17, and DEV1-6,} respectively.
Organization representatives were compensated with a \$25 Amazon gift card\footnote{Participants were offered a choice between a Canadian or American Amazon gift card.} for an hour-long interview, and sex \add{industry} workers were compensated with a \$40 Amazon gift card for a 30-minute interview.\footnote{Sex work is both high risk and can be relatively highly paid, and thus we compensate participation in line with past work on sex work~\cite{elissa1}.}
\add{Participants were assured that their names and organizations would not be reported in our findings. One way that we preserved anonymity was by awarding the gift card code before ending the interview to avoid communication over email.}
This study was approved by the appropriate institutional ethics review boards.

\subsection{Position Statement}
\label{position_statement}
\minoradd{
The authors' position on the content presented in this study evolved over the course of this study.
This evolution is an example of the importance of the wider problem of discussing ethics in technology.
The authors of this paper identify as scholars studying technology and the sex industry.
This study began as an optimization question of how we can improve technology-facilitated outreach to victims of sex trafficking.
The addition of researchers to our team who had a broader grasp of current multi-disciplinary research literature related to sex work and sex trafficking was a turning point in the framing of this study.
Through a review of research, resources, events, and sex-worker-produced content, our position changed from studying how to optimize technology that would better address the sex trafficking problem to understanding the process of technology-facilitated outreach and experiences of all stakeholders involved in this process.}

\minoradd{
To summarize, the authors of this paper prioritize sex industry workers' input in line with participatory action and research justice frameworks~\cite{research_justice}.
As aforementioned, after a more thorough review of the literature and conducting the initial steps of the study, the authors of this paper came to hold the position that the legislative approach that best protects adult sex workers is to refrain from arresting them for sex work-related crimes.}
\footnote{This position is in line with the consensus on harm-reducing legislative approaches in academic literature and suggested best practice by organizations such as the World Health Organization, ACLU, and Amnesty International \cite{who,aclu,amnesty,sexhum,elissa1,kempadoo1,kempadoo2,revolting_prostitutes,playing_the_whore,kenway,decrim}.}
\minoradd{The evolution in positionality also affected our interview, analysis, and paper writing process: At the time of the interviews, we prioritized the viewpoint of each respective interview participant; thus, participants may have felt their own view point reflected back by the interviewer. However, by the time we analyzed the data, all research team members held the aforementioned anti-incarceration position and thus the positions expressed by some interview participants may not have been represented on the research team during the analysis process.}
%\footnote{Research suggests that criminalization of sex work and sex industry workers' clients \cite{sexhum,assistant1,assistant2} (1) harms victims of exploitation and trafficking by making it riskier for them to ask for help from law enforcement or anyone else and discouraging clients from reporting trafficking \cite{sexhum,FOSTA_musto,decrim,internet_sex_work} and (2) prevents sex industry workers from protecting themselves, helping or working together with other sex industry workers, reporting incidents to law enforcement, or sufficiently screening clients \cite{hoefinger,decrim,internet_sex_work}.}

\minoradd{
Though the researchers identify with varying genders, sexual orientations, races, ethnicities, and ages, we recognize the limitation that we do not necessarily mirror the full range of identities held by our interviewees.
We acknowledge the marginalized and often precarious positions of sex industry workers, and also our complex position as researchers, and we follow suggested research justice principles in the design of our study to the best of our ability \cite{research_justice}.}

%It further .
%Decriminalization would also reduce the space for law enforcement to enact violence on sex workers \cite{hoefinger,sexhum,decrim}.
%It should also be noted that law enforcement, as with other ``survival crimes,'' disproportionately targets marginalized individuals such as transgender people or people of color trading sex \cite{walking_while_trans_repeal}.

%Interviews with sex \add{industry} workers solidified our viewpoint that there was more to the sex \add{industry} than what was understood by the organizations interviewed alone.\footnote{Mainstream anti-trafficking reports that conflate sex work and sex trafficking \cite{who_buys_sex,thorn_report,glotip,appg} are widely cited among technologists, the rescue industry, and the media, where they \add{contribute to} moral panic. Sex \add{industry} workers' weariness from being misunderstood by researchers and the media has led to guides for sex \add{industry} worker rights activists and media representatives on how to talk about sex work and dispel common myths \cite{swan_guide,open_letter,media_guide}.}

% ``Nothing about us without us'' was a movement originally pushing the inclusion of voices of people with disabilities when designing legislation regarding disabilities \cite{nothing_about_us}; the movement has since been adapted onto other areas including sex work to push for the inclusion of sex workers when designing policies, implementing programs, or studying sex workers \cite{red_umbrella_nothing_about_us}.

\subsection{Limitations}
Each step of our method has limitations.
We required all interviews to be with participants over 18 years old, which means that we cannot evaluate anti-trafficking campaigns targeted at minors.\footnote{\add{By law in the United States, all minors encountered by \sas\ are referred to the National Center for Missing and Exploited Children, not service organizations, making minors a special case not included in our study. Communication and outreach technologies for minor abuse warrants further study, building on existing CSCW research on the algorithms used to mitigate sexual abuse of minors~\cite{10.1145/3479609}.}}
This study was also Western-centric and focused on the sex \add{industry} in North America.
Another limitation was that all communication was done in English and recruitment of sex \add{industry} workers was limited to those who had access to our digital flyers.
While we addressed certain limitations to the best of our ability (e.g., requesting many organizations to send out flyers, including organizations we did not interview, using social media in such a way that sex \add{industry} workers could re-share the flyers, and making it as easy as possible to sign up anonymously) we acknowledge that these efforts still leave out those who do not have internet access, or who do not have personal phones,\footnote{However, those without digital contact are largely outside the scope of this research as they are unlikely to be reached through technology-based outreach~\cite{unreachable} which relies on being able to digitally contact those whose advertisements are scraped.} and those who are not involved \add{(or no longer involved)} in the sex \add{industry} worker or rescue organization communities.
Despite these limitations, the population of \add{sex industry workers} that we recruited is diverse in multiple senses: a variety of experience in sex work from outdoor or street-based work to escorting to digital-based work, a variety in genders and sexual orientations, and a variety in races and ethnicities.
Because our interview protocol covered a wide range of experiences in the sex industry, we tried to avoid topics or questions that we felt might be irrelevant or re-traumatizing for a given participant, and, of course, allowed them to skip any questions that made them uncomfortable.

%\subsubsection{Scammers}
%Due to the open yet nearly anonymous nature of our recruitment, along with the gift card compensation promised, we encountered scammers: people who claimed to have experience with sex work in North America, but clearly did not.
%These people often signed up for multiple interview slots at once, using different email addresses.
%We employed evolving protocols to end interviews with scammers and remove their data from our study.
%We also used protocols to cancel interviews during the signup process if it was clear they were scammers, since the calendar would otherwise be filled with scammers.
%We do not reveal our protocols here so that they can be used in future studies; even during the study, our protocols had to evolve due to the adversarial nature of detecting scammers who were learning our protocols.
%Generally, we tested participants' basic knowledge of sex work in North America without raising the suspicion of the scammer that we were testing them.

\section{Findings: Tensions around Unsolicited Messaging}

\add{We found that a mismatch of goals between \sas, its client organizations, and the recipients of outreach led to an implementation of a scalable unsolicited messaging system that both increased the reach to those who might need support and also led to a significant volume of spam.} 

Outreach messaging via \fs\ begins with a single SMS message mass-texted to a batch of phone numbers scraped from online sex work advertisements.
\add{Organization representatives who use the platform told us that prior} to the availability of technologies like \fs, organizations manually went through online advertisements and that technologies like \fs\ allowed them to send messages at a far greater scale.
\add{Organization representatives found this new batch method compelling because it allowed them to reach more people in a targeted way, which saved time:}
\begin{displayquote}
``We can reach potentially hundreds of women with a few buttons. Yeah, [manually visiting sex ad sites] it was taking volunteers 2 hours to reach 25, and out of those 25 we may have 1 or none responses back. But now with this platform, we can geographically target who we want to target, how we want to target them, and track it all on one thing. It's just very organized and it's going to save us a lot of time.'' (ORG6)
\end{displayquote}

\add{While organization representatives acknowledged that this increased capacity also increased the volume of unsolicited messages (or spam) they sent, they did not perceive this to be harmful. 
However, in our interviews with sex industry workers, we learned that this increased volume of text messages presents potential threats.
We also learned about the} reliability and trustworthiness of SMS messages as compared to other platforms as a vehicle for outreach \add{to marginalized populations such as} the sex worker community. 

\add{In the following sections, we describe in detail the organizations' perceptions of the efficacy of their communication systems and contrast these perceptions with the experiences of the sex industry workers whom we interviewed.}
% \subsection{Why Outreach?}
% \subsection{What is Outreach?}
\subsection{Why Send Unsolicited Messages?}
% \subsection{Goals of Outreach Technology}
\label{goals}

\add{\sas\ and organization representatives had varying goals for outreach. But in all cases their mass SMS outreach received little response.
\minoradd{``Average response rates'' (ORG7), defined as any positive or negative response to a message on \fs, tended to hover around 5\%, with some variation between organizations.
As stated by ORG14, ``if you are a numbers oriented, goal oriented person that defines success by that, this won't be for you.''
Organization representatives seemed largely okay with the low response rates.}
% \begin{displayquote}
% ``I['d say our] average response rate is 5\%.'' (ORG7)
% \end{displayquote}
}

\add{By contrast,} \sas's overall stated mission includes a vision where ``no one is exploited for labor or sex.''
% \elissa{I would suggest not adding the below, I think it will add to the tone issue toward the organizations}
% \add{This mission envisioned that with scale, the technology would, in fact, reach more people.
% They saw the problem as one of reach volume and accuracy and not necessarily substance.} 
\sas's technology business functions by providing data -- mainly sex \add{industry} worker phone numbers scraped from online advertisements -- to frontline service provision organizations.
The developers we interviewed said the goal of their online platform is to ``effectively connect advocates to potential victims'' (DEV3).
Specifically, \sas\ representatives involved in goal-setting stated that the goal of building this technology is ``to facilitate helping people out of the sex trade'' (DEV1), i.e., to enable organizations to facilitate workers stopping sex work \add{regardless of where they fall on the autonomy continuum}.

\subsubsection{\add{Intentions of those performing outreach}}
\add{We observe misalignment between outreach organization intentions and the needs of those they reached through their communication systems.}
\sas\ is secular, but most of their clients are religious organizations.\footnote{Religion and anti-trafficking campaigns have a history of going hand-in-hand \cite{zimmerman_religion}. Additionally, prior work on the various ways that technology can aid religion and proselytism posited that ``NGOs will use data to profile their supporters and then target messages at specific groups'' \cite{religion_tech}.}
\sas\ requires client organizations to offer concrete services (or have strong referral programs to provide material services), not just prayer, and to be vetted by an existing \sas\ client.

We found that organizations' immediate short-term outreach goals tended to correspond to the size of the organization.

\add{Smaller and less-established organizations} tended to be more faith-based and reported goals centered around relationships: they wanted to ``walk with'' sex \add{industry} workers on their ``journey,'' gain sex \add{industry} workers' trust, and let sex \add{industry} workers know that they are there if they need them.
\minoradd{For example, ORG16 considered themselves successful in reaching their goals if:
``we were there even if nobody else was. So they know that we will be consistent. So that's successful to me.''}
Central to religious organizations' goals was ``to let the women know that they're loved not just by us, but by God'' (ORG2).
These organizations reported that \add{their clients} found this religious aspect to be important, and they told \sas\ that ``prayer request'' was often requested by sex \add{industry} workers; however, \add{when \sas\ investigated text message conversation outcomes, they found that very few, if any, sex industry workers requested prayer.}
% \sas\ also reports that deeply religious organizations tend to be more personally invested in their outreach efforts.
% One organization changed their outreach efforts from secular to religious because they believed it helped show the sex workers that they are loved.
\add{Relatedly, these organizations typically offered referrals to religiously-affiliated direct services (e.g., healthcare, shelter) which created discomfort for some support recipients.}%While they did offer services or referrals---even if those came with religious connotations---prayer and love were seen as primary services offered, on par with healthcare and shelter.

\add{Bigger and more well-established organizations} were also often faith-based. \add{However, these organizations} focused their work more on case management and providing concrete services and referrals such as shelter, financial assistance, legal assistance, etc.
They often said they sought to meet whatever needs the client had:
\begin{displayquote}
``It's just a lot of different, whatever that need is, we do try to meet it. That could be housing. Sometimes I could get clients that are still in the life, but they are still staying connected. So if they stop by the agency [...]
% I tell them `hey if you have ever hungry or if you need a change of clothes let me know. I am at the agency. This is where I'm at. You can always ask the front desk if I am there.' I leave sometimes snacks or things that they can just grab or a blanket or if they need clothes [...] 
If I'm in the building I will come downstairs and I will grab a fresh pair of clothes or whatever and get some snacks or just whatever to try to minimize exploitation if that makes sense so they don't do things to have their needs met.'' (ORG12)
\end{displayquote}

\subsubsection{Ideal outreach goals for outreach recipients}

Almost all sex industry worker participants reported needing support or services at some point, \add{but often not those the organizations (small or large) were offering}.
These resources included financial assistance, legal or technical assistance, healthcare, accounting or tax assistance, and mental health services.

\begin{displayquote}
``Counseling is one of the best things that they can offer. And they could also offer housing because there are some workers who work in the street, and they don't have anywhere to sleep. [...] Also offer financial assistance.'' (SW2)
\end{displayquote}

Most participants reported receiving these resources from colleagues in the sex industry.
Some reported receiving resources from friends and family, clients, or organizations such as those we interviewed, and a few had not had their needs met.
Sex \add{industry} workers we interviewed overwhelmingly preferred services that were offered by other sex \add{industry} workers or organizations led by sex workers.
%\begin{displayquote}
%``I have requested assistance from a sex work mutual aid org.'' (ST3)
%\end{displayquote}
% When discussing reaching out to help other sex workers, sex worker interview participants reported their help being positively received because of their identity as a peer.
% \begin{displayquote}
% ``I work in the clubs, there are some people who also work in the streets, so I tend to meet with them in a daily basis places in the clubs and that is how I get to know them.'' (ST2)
% \end{displayquote}
\begin{displayquote}
``We probably already know of each other's situations, and have a relationship with each other-type of thing. That's what mutual aid is.\footnote{Mutual aid is exchanging of services or care within the community \cite{we_too} that some describe as ``passing around the same \$100 between us'' (\url{https://www.autostraddle.com/how-new-anthology-we-too-essays-on-sex-work-and-survival-honors-sex-workers-truths/}).}
You are both mutually in a similar situation, and you are helping to quell the harm that they’re experiencing or difficulties they're experiencing [...]%'' %(SW12)
%\end{displayquote}
%Ideally, the clients of technology-facilitated outreach could include sex-worker-led organizations, since other outreach from such organizations was more well-received due to their knowledge and experience providing assistance without causing harm.
%\begin{displayquote}
%``
I would prefer to be around individuals that are actually in the industry. [...] There’s always the risk of when an organization is not completely run by individuals in that community, they end up causing harm, rather than mitigating it, without having that direct lived experience of what’s appropriate or helpful or not.'' (SW12)
\end{displayquote}

\begin{figure}[ht]
    \centering
    \includegraphics[width=0.95\linewidth]{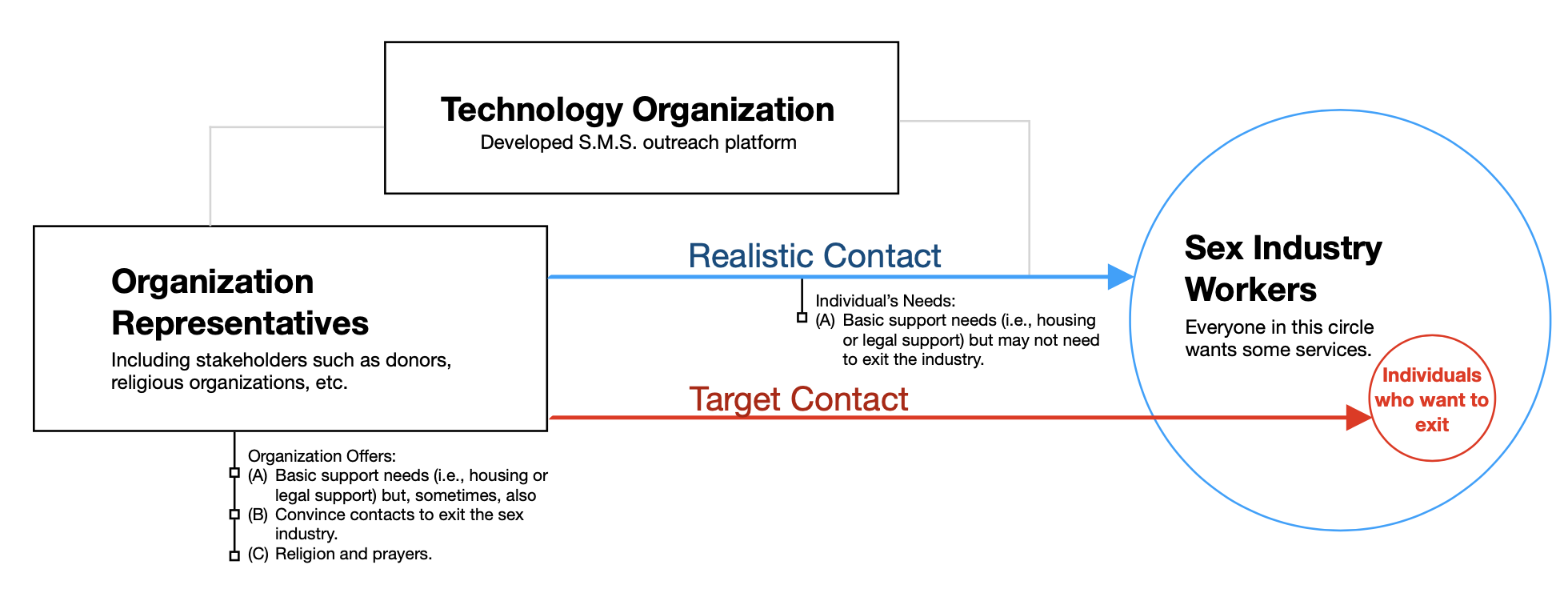}
    \caption{\add{Overview of goals of stakeholders. Organization representatives have various goals including offering support for basic needs, convincing outreach recipients to exit the sex industry, and offering religion or prayer. Through \fs, they contact a large set of sex industry workers, virtually all of which have some basic-support needs, but few of whom are looking to exit the sex industry. \sas\ created the platform that these organization representatives can use to contact the larger set of sex industry workers.}}
    \label{fig:aid_market}
\end{figure}

\add{Figure~\ref{fig:aid_market} depicts an overview of the stakeholders' goals described in this section.}

\subsection{\add{Misalignment around} Proactive Outreach}

\add{In this section, we discuss in depth our findings regarding the misalignment between organization representatives and sex industry workers with regard to the types of outreach (offline and online) they consider beneficial.}
%The organizations we interviewed all conducted proactive outreach, either sending unsolicited SMS messages or visiting sex workers in their workplaces.
Some representatives said they conducted proactive outreach because they were afraid to ``miss people'' (ORG10) by waiting for them to reach out.
\add{This motivation sometimes led to communication insistence, even if a contact notified the organization that they were not in need:}
\begin{displayquote}
``A motivator for all of us [is] when somebody says `where were you?' [...]
If anybody says don't call me again, I may or may not, depending on how hard it is.''
% [...]
% But via the text I would still just say, `if you want to keep my number, you are still welcome to reach out anytime.' ''
(ORG14)
\end{displayquote}

Sex \add{industry} workers varied somewhat in their views on proactive outreach.
A few sex \add{industry} worker participants were or would be grateful for proactive outreach offering support or resources.
\minoradd{For example, SW18 said ``if somebody contacted me about support, that would be great.''}
Some \minoradd{participants} asked us for referrals to resources during the interview \minoradd{and} we responded by giving information about the Sex Workers Outreach Project and their Community Support Line.\footnote{\url{https://swopusa.org/contact/}}
Other \minoradd{participants} pointed out that organizations only need to send unsolicited messages when they are not trusted in the community, and that there was no need for such organizations to reach out:
\begin{displayquote}
``The trafficking-based organizations are the ones that are more likely to reach out---the SW [sex worker] run organizations don't need to because people already know each other and it spreads through words of mouth.
[...]
I mean, there are organizations that go to the place of work, like the Christians who go to strip clubs, but that's a waste of everyone's time. I don't think there is a way to reach out.''
(SW1)
\end{displayquote}
Some sex \add{industry} workers who were skeptical or uncomfortable with outreach expressed concern about \add{the breach in privacy and potential harm that could result from unwanted outreach}.
\minoradd{For example, when asked how they ``feel about them using that information and finding you this way,'' SW7 responded ``I feel that my privacy has been compromised.''}
% \begin{displayquote}
% (Interviewer: ``How do you feel about them using that information and finding you this way?'')
% ``I feel that my privacy has been compromised.'' (SW7)
% \end{displayquote}

% \elissa{would cut this paragraph and leave it for the discussion} Proactive methods may not be viable mechanisms for outreach, given that outreach is viewed by some as unhelpful and potentially harmful and that even those seeking resources might find them online. We also show in section~\ref{platforms} that the communication platforms being used (namely SMS text) may not be an effective form of outreach.

\subsubsection{Reactive Outreach}
\add{In contrast to the misalignment around proactive outreach, all of the organizations we interviewed reported better success with} reactive outreach such as operating a 24-hour hotline or receiving referrals from other organizations or people.
Many felt that offering services that could be accessed when recipients requested them made it more likely that the support would be effective.
\minoradd{As ORG11 said, ``I always try to start by them reaching out to me. Because then they are a lot more open.''}
% \begin{displayquote}
% % ``I think that it is important to give people the option of reaching out. And so just like domestic violence victims, when they are ready. [...] I've gone to trainings where a lot of survivors say, you know it is like, the person has to at least want to be reaching out.'' (ORG12)

% % ``If they want the service they are usually interested in the services right there in the moment.'' (OR16)

% %When it is a referral from outside, I get their information if it is available but% 
% ``I always try to start by them reaching out to me. Because then they are a lot more open.'' (ORG11)
% \end{displayquote}
\subsection{\add{Misalignment around} Outreach Communication Platforms}
\label{platforms}

\add{Our interviews also identified limitations of the communication system related to reliability of the text platform for outreach, the trustworthiness of the format, and associated safety risks.}

The organizations we interviewed proactively found and reached out to sex \add{industry}  workers through a combination of technology-facilitated outreach (e.g., using \fs\ or manually using sex ad websites to contact sex \add{industry} workers) and in-person outreach (e.g., going to strip clubs and massage parlors). 11 organizations (out of 14) reached out to sex \add{industry} workers via contact information in sex \add{industry} workers' online advertisements, and 10 (out of 14) visited sex \add{industry} workers in the locations where they work.

% \elissa{I think we could cut this paragraph as it's well summarized in the prior one} Beyond in-person outreach, organizations primarily performed outreach over SMS text messaging, including via \fs.
% In addition to using \fs, at least one organization manually visited online sexual advertisements and contacted sex \add{industry} workers through advertised means (mostly phone calls).
% In addition to SMS texting, especially with referrals from law enforcement, crisis hotlines, or other organizations, representatives chose to use any platform that is suitable for the situation, including video calls, phone calls, or email.

\subsubsection{SMS text messaging is not reliable as a primary vehicle for outreach.}

When communicating via \fs, organizations found that ``phone numbers change constantly,'' (ORG10) or that the phone numbers ``switch hands quickly'' (ORG7) away from the person with whom they were originally talking.
Representatives we interviewed made it clear that they could not speculate on why the phone number was taken from the person they were communicating with, but guessed that the phone was taken by a trafficker. 

On the other hand, the sex \add{industry} workers we interviewed mentioned that their number changed either because they shared their device with others or because they used a proxy texting service. 
\begin{displayquote}
``I only give my phone number if it's someone that I have met before. Sometimes I'll do an online phone number through \add{[online texting service]}. %TextNow.
And then I have a burner email address.'' (SW1)

``I would leave my \add{[online texting service]} %TextNow
number.'' (SW15)
\end{displayquote}
\add{This online texting service} assigns a phone number to the subscriber for the length of the subscription; after the subscription ends, the number is recycled to a new subscriber.

Beyond our own results, reports from sex-worker-led organizations also note that sex \add{industry} workers often share phones among themselves, and some do not have stable access to a phone~\cite{unreachable}.
%Organizations knew that sex workers sometimes ran out of ``minutes'' to continue using their phones while they were in contact with organizations. %
\add{To remedy this, one} organization \add{we interviewed} purchased temporary phones for their clients while they were working with them\add{, though this does not resolve the issue of making contact in the first place.}

Unlike messaging via other platforms, text messaging does not offer any identifying details about the sender or recipient. As ORG1 put it, ``we don't know who we're talking to.'' Organization representatives were apprehensive of reaching a ``pimp''\footnote{We note that the word ``pimp'' is gendered and racialized, and not preferred by sex \add{industry} workers \cite{stella_info_sheet}.} when performing outreach via \fs.
\begin{displayquote}
``Because it's a text, we don't know, like, if it's a pimp receiving or if it's a worker.''
% We just trust that it's going out to somewhere that could be needing it.''
(ORG3)

``Yeah, safety wise, I mean, we can't control if a trafficker or abuser or whoever gets our message, you know, I mean, the risks, that risk is still there.'' (ORG6)
\end{displayquote}
Some sex \add{industry} workers did work with others to help them with scheduling, screening, and communication, and others aspired to someday hire an assistant to take care of time-consuming administrative tasks.
\begin{displayquote}
``I do have someone who is kind of helping me out. And I will say, it has really worked out for me. I was in this area because right now I tend to get more clients as compared to before, I was on my own.'' (SW2)

``I do everything myself. I think it would be nice to, maybe, one day have an assistant to handle the communication, because it’s a lot of work.'' (SW23)
\end{displayquote}
It is likely that some outreach via \fs\ reached these assistants rather than the sex \add{industry}  workers.%, though they are very unlikely to be traffickers or ``pimps''\footnote{The law prohibits sex workers from using any business services at all; services such as drivers, administrative or marketing assistance, and security are characterized as ``pimping,'' ``trafficking,'' or other exploitation \cite{assistant1,assistant2}.}.

% \begin{displayquote}
% ``I know a lot of times the women will call a pimp a boyfriend.'' (OR14)
% \end{displayquote}
%\paragraph{Sex workers needed to vet organizations}

\subsubsection{SMS messaging may violate communication preferences.}

Sex \add{industry} workers we interviewed often had rules for contacting them such as work hours, only contacting through a particular communication mode (e.g., email), and other ways of establishing their digital boundaries; \add{prior work finds that these boundaries are one of seven forms of sex-worker digital safety}~\cite{elissa2}. 
\begin{displayquote}
``My advertisements clearly state my work hours, but an organization reached out outside of those, which was very frustrating.'' (SW24)\footnote{This quote is from the one interview where notes were taken instead of recording due to a technical error and thus may be a slight paraphrase.}

``I'm glad that no one got my phone number and I didn't get any advertisement or any offer through my phone number.'' (SW20)
\end{displayquote}
Only one organization was aware of these preferences, or thought to mention them, but wasn't sure why someone would want to take these precautions.
\add{This provides additional support for the concept of involving experienced sex industry workers, who would be aware of these issues and their reasons, in developing outreach methods.}
\begin{displayquote}
``Some ads say `no text.' Very few but some. [...]
I don't know why they would have it, if it really is a data plan issue, or I always remember [the story of] that woman who said sound will set off the surveillance.'' (ORG14)
\end{displayquote}

\subsection{\add{Tailoring Outreach to Sex Industry Workers' Identities}}
\label{identity}

\sas, sex \add{industry} workers, and organizations interviewed did all agree that identity is an important aspect of the help offered.
For example, they agreed that certain populations like LGBTQ+ need specialized medical care, or experience homelessness more often, and that resources for these populations should be tailored.
\begin{displayquote}
``There are also really high rates of homelessness among LGBTQ+ youth. And so one example of how we could tailor a message is [...] in my initial message I can talk about potential homelessness or the unique concerns of being this kind of minority exploited group in this space.'' (DEV6)
\end{displayquote}
% However, sex workers are wary of their demographic information being used to direct outreach.

\fs\ covers more gender identities and types of sex work than its platform predecessor, and allows organization representatives to filter the phone numbers to which they are sending an initial message by gender, age, location, and the time the sex ad was last updated.
\sas's goals in implementing this functionality are to help organizations tailor their outreach and to measure effectiveness for different identities.
However, such a filter allows some people to be left out of services offered,
especially considering the error rate with collecting location and age (due to sex \add{industry} worker cybersecurity or business practices such as creating multiple personas).%: the question of whether such filters were in line with their goals is complex.

Some organizations \add{we interviewed} only catered to certain demographics.
For example, several organizations only helped women, a few organizations specialized in the LGBTQ+ population, and a few organizations targeted a certain age range.
Before they had demographic filtering in \fs, organizations would often encounter sex \add{industry} workers outside their target population; when this happened, some organizations simply cut off contact, and other organizations referred the sex \add{industry} worker to an organization who did cater to them.
%Now, with demographic filtering those who would have been helped by the latter set of organizations will now not be contacted.
With the demographic filtering enabled by the system we study, those who would have been contacted by the former set of organizations will now not be negatively affected by an organization reaching out and then having them ``stop the conversation'' (ORG7) when they find out that they are not in the organization's target demographic.

A few religious organizations also used \fs's gender filter to ``protect the husbands'' (ORG1) by ensuring that male representatives are not tempted by female sex \add{industry} workers when doing outreach.

Sex \add{industry} workers similarly acknowledged that because some services were specifically needed by certain demographics, members of those demographics wanted organizations to reach out specifically to them, so long as it was not harmful in other ways.
\begin{displayquote}
``I think, obviously, having medical care that is gender affirming, for example, and that is not violent in all the ways.'' (SW16)

``Because then, they could filter the kind of support they are offering and offer something specific.'' (SW21)
\end{displayquote}
Some sex \add{industry} workers who were women were ``naturally suspicious of men and straight people'' (SW13) and some who were men explained that ``No offense, I relate more to men'' (SW17).
\begin{displayquote}
``I find that, in general, I appreciate when people I’m seeking support from identify somewhere in the queer community. If they identify with sex work, that would be really cool, too. Or like women, I’m more comfortable with women. But straight males, I’d be hesitant.'' (SW22)
\end{displayquote}
Some people of color preferred outreach to only be from people of color.
\begin{displayquote}
``For something like this that is so personal, it's like therapy. I would need my therapist to understand what racism is first-hand, not because you heard it from CNN news, it's that you experienced it on a regular basis. Yes. When it comes to my sex work, I would definitely need somebody of color speaking to me about these things.'' (SW17)
\end{displayquote}

However, members of the BIPOC and LGBTQ+ communities expressed concern about people knowing these aspects of their identity before they divulged it themselves.
\minoradd{SW1 said, ``I don't think they should know information about your identity if they are going to reach out to you.''}
% \begin{displayquote}
% ``I don't think they should know information about your identity if they are going to reach out to you.'' (SW1)
% \end{displayquote}
We know from prior work that taking users' personal information and using it to unexpectedly ``hyper-personalize'' online advertising results in ``highly negative reactions'' \cite{hyper_personalize}, however, additional concern stemmed from a fear that if an organization knows personal information, then they might discriminate based on it or, even worse, \add{any racist or homophobic law enforcement officers} could also find their information.

\add{\section{Findings: Tensions around Outreach Message Content}}

Beyond initial contact, outreach messages sent by organizations were highly relevant to the impact on sex \add{industry} workers and the efficacy of the solution for all three stakeholders.
As mentioned, response rates to initial mass-text outreach messages were very low.
\add{But} even when someone responded, several aspects of the continuing conversation influenced whether sex \add{industry} workers remained engaged and whether agencies achieved their goals.
Specifically, sex \add{industry} workers and organization representatives had \add{differences} regarding continuing communication after the recipient had not expressed interest, the nature of the communication, and risks to workers from continuing communication.
\subsection{Unsolicited Outreach towards the Goal of ``Exiting Sex Work''}

As communication continues after initial contact, the goals of each party become clearer (see Section~\ref{goals}) along with their conflicts.
Some organizations continued proactive outreach after the recipient initially expressed a lack of interest.
The organizational participants who did this felt that their recipients required assistance without knowing or wanting to express it.
Specifically, they believed that sex \add{industry} workers are working under ``false autonomy'' (ORG3) or harbored the long-term objective of having sex \add{industry} workers ``exit the life'' (stop doing sex work).
Their technological outreach strategies aligned with these organizations' websites, which stated that ``exit'' or ``abolition of the sex trade'' was their ultimate mission in ``fighting sex trafficking.''

Sex \add{industry} workers interviewed, even those who would be grateful for outreach, specified that outreach should be offering support or resources and not ``disguised as some type of anti-trafficking effort'' (SW12) or ``in the vein of an anti-trafficking sentiment that saw all sex work as the same form of exploitation'' (SW1).
Sex \add{industry} workers told us they rejected being told that what they are doing is wrong:
\begin{displayquote}
``I knew I was doing the right thing and no one was supposed to tell me what I should or shouldn't do.'' (SW10)

``There are certain organizations that sort of want people to quit sex working so that they can be assisted. Personally, that doesn't sit right with me.'' (SW9)
\end{displayquote}

Specifically, the sex \add{industry} workers we interviewed explained that the incentives for doing sex \add{industry} work were very similar to why they might choose any job: income.
We interviewed sex \add{industry} workers who ``didn't have a[nother] stable job'' (SW19), who did sex \add{industry} work to ``earn a living since my mom could not be able to take care of all of us [their family]'' (SW11), or ``for cash to pay for voice lessons'' (SW13).
Others started working in the sex \add{industry}, ``right after joining college [...] to try my best to make ends meet'' (SW6). Similarly, SW7 shared: 
``Just after joining college, I became broke and I saw it [sex \add{industry} work] as the better option to gain some money.''
% A common sentiment among sex \add{industry} workers was that sex \add{industry} work is relatively more stable, can be more lucrative, and/or gives them more control than ``giving my whole life over to some low wage service sector job'' (SW23).
% Some said sex \add{industry} work was preferable to ``working 80 hour weeks at a vanilla job and still struggling to pay all my bills'' (SW3).
% \begin{displayquote}
% ``It was just really convenient, and I ended up being good at it and joined the industry, because I felt like I had more control when it was difficult to find stable situations.'' (SW12)
% \end{displayquote}
% A few sex \add{industry} workers said they were influenced to do sex \add{industry} work because a friend, partner, or family member(s) was already doing it.
% \begin{displayquote}
% `I was introduced by a friend and since I was at the lowest point of my life I saw it as the better option to make ends meet.'' (SW5)
% \end{displayquote}
These findings align with those from prior work on sex work~\cite{camming_jones} and speak to the continuum of autonomy that exists in the sex \add{industry}.

Even by many organizations' account, the majority of individuals they talked to did not see themselves as victims of trafficking.
In order to achieve their long-term goal of rescuing (or ``exiting'') \add{sex industry workers}, most organizations we interviewed expressed that their job is to accompany sex \add{industry} workers on their journey to the realization that they had been working under ``false autonomy'' (ORG3).
\begin{displayquote}
``A lot of my clients were men trading sex and didn't realize that is out of their autonomy until we have conversations. And it was them realizing it.'' (ORG3)
\end{displayquote}
One organization representative anecdotally recounted a conversation in which they tried to convince a sex \add{industry} worker to stop doing sex work by arguing that their body could not do the work forever---recruiting self-interested and pragmatic arguments for leaving the business over moral ones:
\begin{displayquote}
``A lot of the time with survivors, [...] a lot of their decisions are based off of other peoples' expectations. Or sense of who they were, or who they are. [One in particular] she was trading sex even when she was staying in our shelter. [...] And we're not going to stop people from doing what they're needing to do, but I told her that, you know, you're young, there's a lot of opportunity and also your body can't do this forever.''
(ORG3)
\end{displayquote}
Like several other organizations we spoke with, organization representative ORG3 did not feel that sex \add{industry} workers used their agency to choose sex \add{industry}  work.
According to ORG10, sex \add{industry} work was ``never a choice'' but a means of ``survival'' or a form of exploitation.  
\begin{displayquote}
``Some of the women that were working in prostitution, whether that was by their own choice, which is never a choice, right. It is like the last-ditch effort. But whether there was somebody behind the scenes as a trafficker, exploiter, or it was just, they had no other means to survival.'' (ORG10)
\end{displayquote}

Sex worker scholars have repeatedly debunked the ideas expressed by organization representatives that some are ``experiencing the sex trade'' (DEV3) without agency or that some sex \add{industry} work should be called ``survival sex'' because the workers need money to survive \cite{we_too,stella_info_sheet,revolting_prostitutes}.
Needing money to survive is not unique to sex \add{industry} workers.

Organization representatives reported that many clients are ``halfway in halfway out,'' (ORG12) meaning that the sex \add{industry} worker wanted assistance without wanting to leave the sex \add{industry}.
Representatives also reported that clients whom they have assisted often ``relapse'' (ORG6) - meaning restarting sex-working.
% Organizations reported never ``closing the charts,'' (OR13) which means that they will always be available, regardless of their client's path.

Similar to the ways that spammers cast wide nets in order to reach the small few who interact enough to give the spam campaign a positive net benefit \cite{spam_econ}, organizations sent messages to a broad audience but focused most of their efforts towards the small percentage of sex \add{industry} workers who were looking and able to leave the sex industry.
\minoradd{ORG5 declared that ``Girls leaving the industry is very important to us, and we're going to put money towards that more than anything else.''}
\subsection{Deterrents to Sex \add{Industry} Worker Interaction: Ideology and Saviorship}

Sex \add{industry} workers were sensitive to the ideology presented in the outreach messages they received.
While many organizations reaching out to sex \add{industry} workers had the idea that they were offering a path for redemption or rescue for the sex \add{industry} workers they encountered, many sex \add{industry} workers we spoke with felt their work was just a job or method to make money.
% These conceptions influenced whether those who were reached out to would reply. 

\minoradd{Sex industry workers interviewed across the spectrum of autonomy displayed an aversion to messages and organizations that included religion in their messaging or services, or which otherwise passed moral judgment such as
conveying that they were looking down upon sex \add{industry} workers or sex work, or conflating sex work and trafficking.
A common complaint was ``I just don't appreciate some of the negative remarks people make about sex workers while extending their help'' (SW5).
SW4 emphasized that people performing outreach ``should avoid stressing that sex work is evil or sinful. They also shouldn't criticize one's source of income.''}
% \begin{displayquote}
% ``I just don't appreciate some of the negative remarks people make about sex workers while extending their help.'' (SW5)
%
% ``They should avoid stressing that sex work is evil or sinful. They also shouldn't criticize one's source of income.'' (SW4)
%
% % ``I think if it’s framed as a `hey we are here to help you, this is what we can offer,' and not `we are here to save you, you are trapped and abolition of sex work is the only way.' '' (ST3)
%
% % ``I feel like the least I should expect from them [is not to] call us prostitutes\footnote{The word ``prostitute'' has negative connotations and is not preferred by sex workers \cite{stella_info_sheet}.}.'' (ST19)
%
% ``You might have someone who might want to help you but they don't treat you with dignity because you are a sex worker.'' (SW2)
%
% %``I would be way less likely [to reply to a religious organization] but if it was a non religious one that also seemed as if it was in the vein of, if it had this idea of sex work that doesn't really make sense, then I probably would have the same idea.'' (ST1)
% \end{displayquote}

\minoradd{SW24 said that they would avoid ``someone who might want to help you but they don't treat you with dignity because you are a sex worker.''}
Outreach recipients \minoradd{did not appreciate} what the messages they received implied about them.
\minoradd{SW24 recalled previous outreach, ``she said she could help me get off drugs---I get that she was trying to be helpful but sounded ignorant because I'm not a person on drugs.''}
% \begin{displayquote}
% ``She said she could help me get off drugs---I get that she was trying to be helpful but sounded ignorant because I'm not a person on drugs.'' (SW24)
% \end{displayquote}

Even those we interviewed who identified as survivors of trafficking found it more helpful if an organization viewed them as having agency.
\minoradd{SW14, who identified as a survivor, described an organization which ``tries to empower our survivors, and that was something that was really helpful.''}
% \begin{displayquote}
% ``[Organization] tries to empower our survivors, and that was something that was really helpful [for me as a survivor].''
% % We had this woman who came in, and we would just talk about different things [...] Just some of the different projects we worked on really helped me see myself in a different light.
% (SW14)
% \end{displayquote}

When vetting outreach organizations, the sex \add{industry} workers that we interviewed checked for religious affiliation: they almost unanimously preferred outreach and services to be devoid of religion (the few who didn't had no opinion on the matter).
\begin{displayquote}
``Religion and sex work do not go together. I feel like a lot of the stigma comes because of religion. [...] It has to be a neutral ground. If it is tied to a church, I don’t know if I really would want to.'' (SW17)
\end{displayquote}
This was often due to past negative interactions with, or perceptions of, religious organizations.
Even those who accepted religion at first because it came with needed services eventually found the religious aspect to be difficult to cope with.
\minoradd{SW4 described their experience receiving needed resources with religion, ``at first it was welcoming but then I started feeling guilty about what I was doing and it affected me mentally.''}
% \begin{displayquote}
% ``At first it was welcoming but then I started feeling guilty about what I was doing and it affected me mentally.'' (SW4)
% \end{displayquote}

Religious organizations interviewed were unaware of the perception that religion was relentlessly touted in their outreach;
several told us that they identified themselves as faith-based when doing outreach to sex \add{industry} workers or that they offered prayer, but insisted that they ``do not proselytize'' (ORG3).
\begin{displayquote}
``We're offering support, but we're not pushing it on anybody. But when people come to us, like, they know we are a ministry. Now, we're not going to force feed them anything, we're not going to charge or force them to partake in a Bible study. We have no, like, agreement that they have to sign.'' (ORG5)
\end{displayquote}
Organizations said they do not push religion yet did not seem to appreciate that presenting as religious (e.g., offering services and a prayer) is not perceived as neutral.
For example, some organizations began outreach conversations by sending Bible verses and introducing themselves as a religious organization:
\begin{displayquote}
``We just say [we are] a group of women of faith, but we always also ask them, do they want prayer? We take scripture. [...] In the beginning, we always send them the scripture.'' (ORG7)
\end{displayquote}
Asking if they want prayer (rather than preemptively giving it) was their way of avoiding proselytism.
Indeed, according to sex \add{industry} workers, avoiding proselytism is necessary but not nearly sufficient.
\begin{displayquote}
``I would feel more comfortable if it was a non religious organization.'' (SW2)

(Interviewer: ``Do you think there are any other characteristics that, maybe, organizations that you wouldn't be comfortable with have?'')
``Religious.'' (SW23)
\end{displayquote}
\subsection{Dangers of \add{Organization Representatives} Texting Sex \add{Industry} Workers}
Because of the criminalized and stigmatized nature of sex work, messages to sex \add{industry} workers could be harmful to those workers because of 1) the risk of non-private message content or notifications giving away the recipient's identity as a sex \add{industry} worker and 2) the risk of law enforcement finding evidence of sex work through surveillance of communication messages, the database of sex \add{industry} worker phone numbers, or untrustworthy organization representatives.

Many sex \add{industry} workers we interviewed were concerned about messages revealing that they were sex workers:
\begin{displayquote}
``Girls who I've worked with up in the massage parlors say things like `If my boyfriend knew I did this, he would kill me.' '' (SW1)

``No one really wants to expose themselves as fssw [full-service sex worker] if possible'' (SW3).
\end{displayquote}
Due to the stigma of sex work, many sex \add{industry} workers are not ``out'': the worker's friends and family do not know that they are a sex \add{industry} worker~\cite{elissa1,elissa2,camming_jones,internet_sex_work}.
In addition, messages related to sex work, including those indicating that the recipient is a sex worker, are not allowed on some online platforms and may cause the sex \add{industry} worker to be banned from the platform~\cite{hacking_hustling_erased,elissa1}.

These dangers described by sex \add{industry} workers were at best a secondary concern to the organization representatives interviewed.
Instead, organizations were primarily focused on carefully designing their outreach messages and conversational protocols to ensure that they sounded human (not like automated spam), built trust, and increased the chance of a positive response to their outreach (the recipient accepting services).
Many interviewees said that they built trust by sending multiple messages over time (``walking with'' sex \add{industry} workers on their ``journey''), sending pictures, or setting up phone calls to prove that they were human and willing to help, not realizing that repeated outreach \add{yields repeated chances for harm}. 
\begin{displayquote}
``Building a relationship and trust is a huge thing. Sometimes it took us to 6 months before the other side say, `hey, come up.' '' (ORG7)

``I know it takes, like, usually like month, 3, 4, 5, when the girls receiving the text message again, then maybe they will respond.'' (ORG5)

% ``We've had we've had a couple people ask us for pictures.'' (OR2)

``I don't expect always to have significant interactions over texts, so I offer a phone call [...] She [one client] said once you send a text 20 times, then she would call. Then she would know that it's real.'' (ORG3)

% ``The system doesn't allow us to video chat or call them directly. So we have to use another number to do that.'' (OR1)
\end{displayquote}
While some organizations said they kept their messages (at least the initial messages) vague enough to avoid the ``assumption'' that the recipient was a sex \add{industry} worker, they nevertheless continued to offer prayer and sentiments that \add{are deemed by sex industry workers we spoke with to be} suspicious.
There were other ways that these messages put sex \add{industry} workers in danger. Some organizations require names for certain services (such as paying bills) which sex \add{industry} workers told us was a deterrent \add{because of the risk of} their information reaching law enforcement.

\label{vetting}
Sex \add{industry} workers also had safety concerns regarding unknown SMS messages.
Just as sex \add{industry} workers described having or aspiring to have a vetting process for their clients,\footnote{As with prior work involving sex workers' vetting \cite{elissa1}, we do not disclose participants' vetting processes here for participant protection.} sex \add{industry} workers also vetted organizations upon receiving unrecognized outreach to ensure that the organization provided legitimate assistance and was not affiliated with law enforcement.
\begin{displayquote}
``I tend to decline several of the services that most of the individuals try to offer me. [...] First of all I have to like look up if the organization, well, some of the people that have been offered help by the organization, if the help did really have a great impact from them. From there I can be able to make like an informed decision.'' (SW2)

% ``I went and tried to get more information about the organization, and other reviews from other people who have got the services from them, if they are true and if that organization is trustworthy. That one gained my trust with them. And I was comfortable getting the services because I was in need.'' (ST19)

``I would just need to know if they are affiliated with a legitimate organization, and that they are not involved in law enforcement in any way.'' (SW18)
\end{displayquote}
As with prior anti-trafficking technology such as the ``Be My Protector'' app \cite{be_my_protector}, the fear that law enforcement could be using outreach methods to find sex \add{industry} workers was a barrier to sex \add{industry} worker engagement.
\begin{displayquote}
``How did they get my information? Are they connected to the police at all?'' (SW22)

``It could potentially be law enforcement, or something.'' (SW12)
\end{displayquote}

Even if sex \add{industry} workers vetted that an organization themselves was not directly law enforcement-related, sex \add{industry} workers were wary of law enforcement finding them:
\begin{displayquote}
``They'll be like, `No, we're safe.' And I'm like, `Well, yes, I thought that about my
[expletive] % fucking
dissertation committee too, and then they ended up not being safe.' I also know that they [the organizations] can get hacked. You might be safe, but some
[expletive] % fuck nut
who wants to dox me isn't.'' (SW13)
\end{displayquote}
\add{\section{Reflections and Discussion with Organizations}}

Our findings suggest that \add{unsolicited communication systems such as the one we study are} ineffective at reaching trafficking victims and at facilitating provision of needed services to sex \add{industry} workers.
At best, \add{the unsolicited communication system we study appears} to be a mechanism through which organizations spam sex \add{industry} workers to reach the few interested in interacting with them.
%, at the expense of exacerbating the lived stigma and marginalization of the sex \add{industry}.
At worst, \add{this technology creates} significant risk of harm to a marginalized, stigmatized, and criminalized group through risk of ``outing'' them to those around them or to law enforcement.
\add{Further, the communication system is agnostic to the content that is sent, leaving it vulnerable to becoming a vehicle for harassment and perpetuating stigma.}

If we consider organizations' use of technology to conduct unsolicited outreach through common spam models~\cite{spam_econ}, then organizations are analogous to merchants ``selling'' services, ideology, and exit from the sex industry.
The ``net benefit'' of this system depends on the benefit to the few outreach recipients who receive valuable services or desire to leave the sex \add{industry}, and the amount of interruption, increased stigmatization, and harm caused to the vast majority of message recipients \add{we spoke with} who do not desire such contact. 

\add{Further, viewed through a design justice lens~\cite{design_justice}, which suggests that technologies serving marginalized communities should be designed by those communities, we consider whether community-led initiatives use these same techniques. Our participants indicated that they did not; }%as one participant described  ex-worker-run organizations may be uninterested in this form of outreach because they are familiar with the potential harms; 
as SW1 put it, ``the trafficking-based organizations are the ones that are more likely to reach out---the SW [sex \add{industry} worker] run organizations don't need to.''

\add{We appreciate the extent to which sex trafficking is a sensitive and important issue. Not all might agree on moral, ethical, or even legal grounds about the types of tradeoffs that should be made in this space.
Given the empirical data we have collected and that is available in prior work \cite{sexy_lies,natasha_gordon,natasha_zhang,natasha_davies}, we believe that the tradeoff between benefit and harm from unsolicited outreach to adults working in the sex industry does not merit the use of this technology.
While the organizations we studied do not, for example, engage with minors through \fs, future empirical evidence with these populations may lead to applications for such outreach systems where some might feel that the potential benefit is worth the risk.} 

As technologists, we must critically consider the implications and non-neutrality of building tools for organizations to scale approaches that have little evidence of efficacy and which \add{can} harm those they purport to assist. In the case that technologists choose to pursue the development of unsolicited outreach technologies, we discuss in the remainder of this section our suggestions to reduce the harms of such systems.

\subsection{Follow-up Discussions}
As discussed in Section~\ref{method}, we first interviewed \sas\ and organization representatives, and then we interviewed sex \add{industry} workers.
After completing those interviews, we distilled a set of \add{suggested} ``best practices'' relevant to technologists and organization representatives. We then held follow-up discussions with \sas\ and three organizations who use \fs\ to do outreach \add{during which we presented a summary of our findings along with these best practices, which are presented in the remainder of this section}.

We conducted these follow-up discussions with three organizations that reflect the range of organizations that we originally interviewed, considering a range of emphasis on religion and openness to feedback on their messaging.
%These follow-up discussions began with a presentation of our findings, framed as ``here's what potential clients said would make them more receptive to outreach from organizations,'' with the goal of ``helping organizations to help people.''
%In our follow-up \add{discussions} we avoided confronting or calling out ideological differences or pointing out factual inaccuracies that are fundamental to organizations' beliefs.
%Where appropriate, we did ask ``what it would take'' to adopt the suggestions.
\add{We used these follow-up sessions to  clarify our suggested best practices and give back the insights we generated to some of the stakeholders of our research. As these followups were not conducted with a phenomenological stance that privileged the realities and contexts of our participants, they are not considered interviews within our original method. As none of the organizations with which we initially conducted follow-ups were open to making changes based on our findings, we did not conduct meetings with additional organizations but will provide them with our published paper once available.
}

% \add{While these discussions helped us to refine} our ``best practices'' for technologists and organizations to mitigate potential harms of technology-facilitated outreach to sex industry workers, \add{they did so primarily in the sense that the organizations we spoke with did not expect to change their practices.}
\subsection{Community-driven Best Practices for Technologists} 

\add{Our findings suggest that anti-trafficking technology for unsolicited outreach to sex industry workers is ineffective and likely harmful.
For technologists building and organizations using such technology regardless, our suggestions}
are divided into four categories: safety \& privacy in direct messaging sex \add{industry} workers (Table~\ref{table:security}), language in direct messaging and websites (Table~\ref{table:language}), services provided (Table~\ref{table:services}), and  website design (Table~\ref{table:website}).
\minoradd{The findings are from our interviews unless otherwise indicated with a literature citation, and the suggested best practices follow from the findings, with suggested implementation details inferred from our follow-up discussions with organizations.}
These best practices are not meant to support the current, non-survivor-centred technological approaches to proactive outreach so much as mitigate their harms.

Some of the suggested practices involve preventing abuse of platforms that reach out to \add{sex industry workers}.
These technical suggestions include implementation of features to allow recipients to easily block further contact, moderation of content sent by platform users, outreach message text suggestions, and enforcement of any communication rules presented by sex \add{industry} workers in their advertisements.
We recommend implementing a ``do-not-contact'' list, automatically appending ``Don't want these messages? Reply `STOP' '' to the end of every message, and adding those who reply ``STOP'' to the ``do-not-contact'' list.
The platform could then follow up with anybody who sent ``STOP'' to ask if they would like to report any harassment by an organization representative; this might be tricky as it would require outreach recipients to distinguish the technology platform from the organization's outreach, and understand to whom they are reporting the harassment.
Any conversations that were flagged through this process as abusive would ideally be reviewed by the platform and appropriate penalties applied to the outreach worker or organization.
Simple content moderation such as flagging words that make sex \add{industry} workers feel judged or unsafe \cite{stella_info_sheet} would also go a long way.
Offering suggested messages to organizations as examples that do not make known the recipient's identity as a sex \add{industry} worker can assist newer organizations in conducting outreach safely and appropriately.
Some advertisements request certain work hours, and we recommend parsing those out and enforcing them.
A ``wish list'' could be to include options for messaging platforms such as Twitter direct messaging (if the client organization has an active Twitter presence), but only if other suggestions are implemented to avoid language that would cause deplatforming.

The remaining suggestions require addressing best practices in the platform's terms of use agreements and related ``best practices'' or ``FAQ'' documents.
We also recommend working with client organizations to be more cognizant of the continuum of the sex \add{industry}, reduce personal information collected, avoid any religion that deters sex \add{industry} worker interaction, list all relevant information on websites, and ensure that they understand the intended policy interpretations and potential harms.
Moreover, we recommend immersion in a broader spectrum of the sex industry (see Section~\ref{survivor_bias}), e.g. through involvement in the vibrant community on Twitter.

As marginalized individuals, sex \add{industry} workers require more sensitivity to their data privacy rights \cite{lgbt_privacy}; scraping their information, storing it in a database, and facilitating proactive outreach from the rescue industry would thus seem to be putting them at risk.
Further, while our study is focused on North America, restrictions on scraping such as those in the European Union under the GDPR should be considered.
Additionally, technologists need to address the risk of a subpoena requiring them to supply all scraped information to law enforcement.
Moreover, one could also view (as many sex \add{industry} workers do) the practice of SMS text message outreach as taking time away from or introducing risk to income-generating activity that the sex \add{industry} worker may be involved in at the time of receipt.
Harms from outreach specific to sex \add{industry} workers are greater than inconveniences due to telemarketing or other spam because of the criminalized, stigmatized, vulnerable, or otherwise marginalized nature of sex work.

\begin{table}[ht]
\begin{center}
\begin{tabular}{| p{0.55\textwidth} p{0.42\textwidth} |}
\hline
 \textbf{\minoradd{Findings}} & \textbf{\minoradd{Best Practices}} \\ 
 \hline
  Many people are not ``out'' as sex \add{industry} workers. \begin{itemize} \item{Messages to sex \add{industry} workers must not give away that they are a sex \add{industry} worker. Even a notification can be dangerous.} \item{It is best to follow any advertised ``work hours.''}\end{itemize} & Technologists should enforce work hours and suggest message text. Platform users perceive they are already censoring themselves, even if they include suggestive messages like prayer. \\  
  \hline
 There is a huge fear of law enforcement finding sex \add{industry} workers' information. \begin{itemize} \item{Sex \add{industry} workers are skeptical, even after reassurance that their information will not reach law enforcement.} \item{Requiring a client's real name is a barrier, even if it is confidential.} \item{Telling them where you found their information will help quell the fear that they have been ``outed'' to the police.} \end{itemize} & Technologists should ensure that, even if subpoenaed, sex \add{industry} workers' information will not reach law enforcement. This is a harm that sets this outreach apart from other forms of unsolicited spam. Client organizations are often required to collect personal information, but they often allow pseudonyms except when paying rent or a bill. Platform designers should work with their clients to reduce the information collected. \\
 \hline
 Sex \add{industry} workers do not want others to know their information. \begin{itemize} \item{Reach-outs should not mention if they know the client is LGBTQ+ or BIPOC until the client makes it known.} \item{Reach-outs should wait for the client to express what services they need, even if it seems clear (e.g., that they are using drugs).}\end{itemize} & Other than information collection inherent in the outreach platform, this suggestion \add{should be intuitive in an FAQ document and is} already in use by organizations with established practices. \\
 \hline
 Sex \add{industry} workers have a range of reactions to receiving outreach. \begin{itemize}\item{Some were uncomfortable with being contacted in their workplace or via their online ads. Some were grateful or would be grateful if they were contacted. A keen sense of telling these two apart would avoid some harms of spam.} \item{If someone is uncomfortable, they want to be left alone.}\end{itemize} & Technologists can enforce a do-not-contact list and append to each text message ``Don't want these messages? Reply `STOP'.'' \add{By doing so in the platform itself, technologists can ensure} the ability to \add{opt out and} report harassment, and appropriate penalties could be applied to those who are reported. \\
%  Others continue contact in case they want support later. This is a complex issue that deserves further investigation and potentially well-designed participatory research. \\
\hline
\end{tabular}
\caption{\minoradd{Findings and suggested best practices} regarding safety and privacy with direct messages.}
\label{table:security}
\end{center}
\end{table}

\begin{table}[ht]
\begin{center}
% \rowcolors{1}{}{lightgray}
\begin{tabular}{| p{0.55\textwidth} p{0.42\textwidth} |}
\hline
 \textbf{\minoradd{Findings}} & \textbf{\minoradd{Best Practices}} \\ 
 \hline
 Even with best intentions, religion \add{tends to preclude} sex \add{industry} worker interaction. \begin{itemize}\item{Many sex \add{industry} workers have past negative experiences or trauma with religion, so they avoid religious organizations or feel attacked when religion is mentioned.} \item{Even clients who find it welcoming at first might ultimately feel it is judgmental and harmful to their mental health.}\end{itemize} & For the technology to be truly effective, client organizations must not offer prayer or mention religion\add{, at least in the initial stages of contact}. While we suggest completely removing religion from messaging and websites, this could also be misleading and otherwise harmful for sex \add{industry} workers who will not accept services from religious organizations and would want to know sooner. \\
 \hline
 Using language that sex \add{industry} workers have chosen makes sex \add{industry} workers feel safer. \begin{itemize}\item{All language sent to potential clients must be checked with current sex industry workers.} \item{Words like ``prostitute,'' ``pimp,'' ``John,'' ``victim,'' and ``perpetrator'' are poorly received by many \cite{stella_info_sheet}.}\end{itemize} & We recommend content moderation on message text. We also recommend being informed with the most up-to-date and methodologically sound research, and immersion within the vibrant culture and community that sex \add{industry} workers produce themselves which includes a Twitter community and regular events for political advocacy, sharing stories about sex work, disseminating art and culture, and presenting academic research. \\
 \hline
\end{tabular}
\caption{\minoradd{Findings and suggested best practices} regarding language in direct messages and websites.}
\label{table:language}
\end{center}
\end{table}

\begin{table}[ht]
\begin{center}
% \rowcolors{1}{}{lightgray}
\begin{tabular}{| p{0.55\textwidth} p{0.42\textwidth} |}
\hline
 \textbf{\minoradd{Findings}} & \textbf{\minoradd{Best Practices}} \\  
 \hline
 Speaking to an organization that focuses on people exiting the sex \add{industry} is difficult for sex \add{industry} workers who might see it as a rejection of their work or who are not ready/willing to exit. & We urge technologists and organizations to include space for sex \add{industry} workers %who represent the majority 
 not ready to exit and who need services but who would only contact an organization if it included their situation. \\
 \hline
 \add{Services which are most useful and requested:}
\begin{itemize}
    \item Financial aid (Rent, bills, mortgage)
    \item Healthcare (Non-judgemental gender-affirming healthcare specific to sex \add{industry} workers without risk of law enforcement)
    \item Accounting/taxes (Help fulling tax obligations without reporting anything illegal)
    \item Legal (Copyright assistance, testifying against traffickers, fighting doxxing)
    \item Mental health (Support groups, peer mentors, counseling for sexual violence, substance use, and depression)
\end{itemize} & It is easy to find client organizations who offer most of these services in some capacity, but not specific to sex \add{industry} workers. For example, organizations expressed that they trained medical staff to be conscious of trafficking survivors' trauma, but that a patient should be assumed as ``just a patient'' (ORG3) until they identified themselves as a survivor of trafficking; this illuminates that there is no medical training specific to sex \add{industry} workers who are not survivors of trafficking. \\
 \hline
\end{tabular}
\caption{\minoradd{Findings and suggested best practices} regarding services provided.}
\label{table:services}
\end{center}
\end{table}

\begin{table}[ht]
\begin{center}
% \rowcolors{1}{}{lightgray}
\begin{tabular}{| p{0.55\textwidth} p{0.42\textwidth} |}
\hline
 \textbf{\minoradd{Findings}} & \textbf{\minoradd{Best Practices}} \\ 
 \hline
 On \add{an organization's} website, \add{it is helpful to} list: \begin{itemize}\item{Common services offered (e.g. housing, financial assistance, etc.)} \item{Testimonials of clients who received help} \item{The organization's contact information}\end{itemize} & This is generally \add{straightforward, though} it does not account for the possible constraints of externally hired web developers. That being said, we specifically suggest listing testimonials of clients who are sex \add{industry} workers (in addition to the survivors who are already there); this suggestion \add{may be} more difficult for organizations who believe that all sex work is coerced, and therefore that a testimonial that doesn't end in exiting the sex \add{industry} is not a positive one. \\
 \hline
 An emphasis on trafficking discourages people who identify as sex workers from engaging. \begin{itemize}\item{There can be a section of the website for people who are being trafficked, but if trafficking is the focus, then those who identify as sex workers tend to avoid the organization entirely.} \item{Make it clear to the viewer that the organization does not believe that all sex \add{industry} workers are trafficked.} \item{The website should emphasize harm reduction \cite{harm_reduction} over functionally unrealistic ideologies such as ``abolishing the sex trade'' or ``ending human trafficking,'' which our data showed actively prevented potential clients from engaging.}\end{itemize} & \add{We urge technologists and organizations to make it clear that they know that sex industry workers are not all coerced into sex work. While many organizations are unlikely to heed this suggestion, it is perhaps the most important for any website. Simply adding the phrase ``sex workers'' to the list of people that they are intending to publicly offer services to is insufficient by itself. It is vital that organizations and technologists adjust their view to accept that not all sex industry workers are coerced to genuinely implement this.} \\
 \hline
\end{tabular}
\caption{\minoradd{Findings and suggested best practices} regarding website design.}
\label{table:website}
\end{center}
\end{table}

% \subsection{Funding and Political Neutrality}
\subsubsection{\minoradd{Barriers to Best Practices}}

Technologists and researchers must consider that though we can define best practices, we cannot force people to implement or abide by them.
Some organizations with whom we conducted follow-up interviews were open to suggestions but maintained that certain rules cannot be changed due to underlying funding and organizational motivations.
For example, organizations reported that maintaining \add{an anti-sex industry philosophy or messaging} on their website is often necessary to stimulate donations and maintain funding.
They explained that their websites' main audience is potential donors, and that they aimed to gain sex \add{industry} worker clients' trust through other means.
Sex \add{industry} workers, however, told us that they read an organization's website to determine whether it was trustworthy (see Section~\ref{vetting}).
\sas\ acknowledged the challenges that organizations face trying to reconcile the needs of those they serve and the values of the organizations that fund the work:
\begin{displayquote}
``[This is] a challenge within the nonprofit sector, when you depend on funding from people who don't understand the issue from the perspective who are experiencing it.'' (DEV1)
\end{displayquote}
Moreover, several organizations have funding sources that are explicitly ideologically or politically binding.\footnote{Some government grants in the United States require an organization to define all sex work as trafficking and refrain from advocating for the ``legalization'' of sex work \cite{aplo}.} % That's right, it says legalization
Anti-trafficking organizations, charities, NGOs and business that cater to them in North America have little to no professional evaluation or oversight.
Some organizations themselves are involved politically, supporting laws and people that promote \add{attempting to} end the sex \add{industry through prohibition}.
\sas\ aims to be ``neutral,'' does not currently actively advocate for policies, and is willing to work with a wide range of organizations as clients.
However, the reality is that all of their of customers are more closely aligned with ideologies promoting the \add{eradication} of the sex industry.
% \sas's technology branch is mostly revenue-driven, meaning that it is mostly funded by these client organizations with non-neutral funding, conflicting with \sas's own aspiration to be ``politically neutral.''
% However, in the context of the harms produced by criminalization, the neutral, evidence-based, harm reduction \cite{harm_reduction} position is to advocate for full decriminalization \cite{sexhum,elissa1,kempadoo1,kempadoo2,revolting_prostitutes,playing_the_whore,kenway,decrim}.

% The juxtaposition of the goals of sex workers and organizations revealed that promoting exit from the sex trade is incompatible with aiming to provide assistance to sex workers.
% In order for organizations to make progress towards the fundamental goal of helping more people through outreach, organizations must be willing to switch their public image from focusing on trafficking to focusing on sex work (with a small, separate section on trafficking), and to find donors who support this approach.
% Grassroots non-religious sex-worker-run organizations have funding with fewer restrictions, but they generally have much less funding; the financial barrier to becoming a Freedom Signal user may be contributing to \sas's politically biased client pool.
% Addressing exploitation in the sex trade is more complex and nuanced than sensationalist anti-trafficking rhetoric: it will require the prioritization of sex worker perspectives and agendas.

\subsubsection{\minoradd{Suggested Alternative Approaches}}

\minoradd{While our suggested best practices are to mitigate harm for overall approaches that we do not condone, we} make three suggestions for alternative approaches and future work to address the context studied in this work. First, effort and money spent on developing technology for unsolicited outreach is likely better spent on developing technology that provably helps victims of trafficking.% or on providing direct services to \add{sex industry workers}
For example, technologies could be developed to allow victims without phone access to reach assistance in public venues where they may be present (e.g., public restrooms); organizations could post advertisements on websites dedicated to sexual services (as some already do) to spread awareness of the services they offer to sex industry workers and the fact that they can assist trafficking victims, giving sex industry workers the option to reach out either about services or about a known contact who needs assistance. Second, development of alternative forms of outreach and service provision should be done \textit{in collaboration} with sex industry workers across the full spectrum of autonomy, in line with suggestions for community-driven technology development~\cite{design_justice}. Third, further research \minoradd{similar to} that conducted here is needed to measure the efficacy of \minoradd{the} technological through analog approaches \minoradd{(e.g., posters in transit stations)}~\cite{samarasinghe2007strategising}. Only with this data can we effectively prioritize the strategies that are most effective and have the lowest risk of harm. 

\section{Survivor Bias}
\label{survivor_bias}
%
%Religion, ideology, funding, and knowledge of the sex industry shape organizations' use and efficacy of outreach technology.
% The perception of all sex work as ``trafficked'' or ``coerced'' plays a large role in how this technology is implemented, directing organizations towards problem solving that may, in fact, harm their cause and waste resources. 
% For example, some organizations told us they are seeking to meet sex workers in-person or build relationships with individuals through custom messages to get them to ``exit'' when they know this will ``never happen.''
% Organizations' insistence on prayer ignores the data (both \sas's and ours) that suggests that prayer (or services that require religious commitment) are mostly seen as non-starters.
% These methods may put sex workers at risk with law enforcement, with platforms that ban sex work, or with those near them if they are ``outed.''
% If organizations understood who they were reaching out to, or did research on their populations, it might shape the messages they leave.
% It might lead them to invest in other aspects of outreach---ensuring their website is inclusive of the majority of the population they serve (which it is currently not) and building a social media presence (again, which most do not). 
While we shared sex \add{industry} workers' reports from our interviews with organizations using \fs, 
we were discouraged by their insistence that having survivors on staff provided insight into the experiences of all sex \add{industry} workers.
Their status as ``survivors'' (former sex \add{industry} workers or trafficking victims) presented a narrative that, by definition, left out the majority of the community that was being contacted through \fs\ and other proactive outreach.
\add{These organizations seemed to fall prey to a kind of survivorship bias, basing their perceptions of the sex industry on the very few who remained in contact with them. In effect, these ``survivors'' that organizations spoke of were believed to be outreach success stories because they remained in contact or even got out of sex industry work (possibly even volunteered with their organization) when, more likely, they were a self-selecting few who were seeking to leave and thus continued to message the organization. This minority of sex industry workers, who differed dramatically from those we spoke with, formed the undisputed archetype of the ``survivor'' for these organizations \add{as described in prior work \cite{natasha_gordon,natasha_zhang,natasha_davies}}. In reality, each organization had only encountered, at best, one of these individuals, overlooking all those who did not respond (at all, or for long) to their messaging.}
It is no surprise, then, that we observe significant misalignment between the platform developers and outreach workers composing messages and the majority of those who are reached by it: sex \add{industry} workers who \add{want to remain in the industry and don't view themselves as ``survivors.''}
The growing emphasis in the CSCW community on placing marginalized participants at the center of technology development \cite{elissa2,constanza_chock} is further supported by our findings; involvement of sex \add{industry} workers who do not identify as survivors of trafficking is necessary.
% The sex industry participants we interviewed were articulate and informative---there is simply no excuse for design approaches that exclude their involvement.
It is not our intention to suggest that trafficking is not a serious problem that is worthy of attention, only that technology that addresses it should better reflect the context and nuanced realities of the sex \add{industry}.

\section{Conclusion}
We interviewed 24 \add{sex industry workers}, 17 representatives from organizations that offer assistance to \add{sex industry workers}, and 6 people who developed \fs, which helps organizations perform outreach to \add{sex industry workers}.
We distilled a set of often conflicting themes that arise based on each group's goals.
% Rescue organizations and the sex workers they aim to help may be seen as ``ships in the night'': rather misaligned.
% Our research identified ways that organizations' outreach could be made more effective and less harmful to sex workers while maintaining the resources that sex workers find helpful.

Organizations can provide resources needed by sex industry workers, but their \add{focus on} ``exiting'' the sex industry makes them helpful to very few.
This misalignment between organizations' goals and ideologies and those of the majority of sex \add{industry} workers makes outreach technology effectively spam, though possibly more harmful.
First, none of our \add{sex industry worker} interviewees expressed that organizations should proactively offer to help members of their community to ``exit,'' a goal that is integral to the mission of the organizations we spoke with---particularly among the more religious ones, and ones with exit-mandated funding.
Second, organizations are often unaware of the risks that sex \add{industry} workers face because their orientation to the problem is founded on a narrative of slavery and extreme exploitation which often rejects nuanced understandings of the actual harms sex \add{industry} workers experience through criminalization and stigmatization.
While organizations say that they do not require identifying information to provide services, long-term help is seemingly contingent on prolonged, in-person contact and stable identity.
Lastly, some organizations' lack of understanding about their target clients means that organizations do not always provide compassionate or secure help; they use text messages to offer prayer or religious services, use identifying information, and direct workers to websites that suggest that their help is contingent on commitment to \add{conservative} values and/or religion.
These strategies are demonstrably deterrents to many sex \add{industry} workers seeking help.

Ultimately, many of the organizations we spoke with \add{believe} that most or all of those they reach are being trafficked, and as a result, they imagine a different population than the primary one they reach through proactive outreach technology---both in how they design outreach technology and in how they use it.
When we attained this conclusion, we presented our findings to \sas\ and three additional organizations to discuss the feasibility of our suggestions, but found that they were only superficially willing to address these issues---they ``would not push religion'' but would still offer prayer, and they ``do not judge sex workers'' but still maintain their \add{anti-sex industry} ideologies.
Ultimately, proactive outreach campaigns work much like spam: prioritizing the few who might be interested over the vast majority who might be inconvenienced or otherwise harmed.
% We had misgivings about our decision to conduct these followups for fear that they might, in fact, provide organizations with methods to further conceal their goals, which are not aligned with sex workers' and are detrimental to harm reduction.
% It's hard to see past the goal of providing services as part of a larger project of ending sex work, when those services are contingent on certain commitments.
% It becomes hard to see them as genuinely well-intentioned.
% The representatives of the sex trafficking industry we spoke may have been doing more harm then good and their reluctance to consider sex work on their terms might preclude them from being stewards of this technology.

Thus it is not just the technology itself that we must concern ourselves with, but the ways in which that technology is being used.
We conclude with the guidance that outreach technology needs to be built \add{in consultation} with and/or by sex \add{industry} workers---and not just those whom these organizations have recruited based on adherence to a narrative of trafficking survivorship.
Even when providing much-needed services, when those services are advertised in a potentially harmful way, the organizations and the technology facilitating their work fail to achieve their goals.
\begin{acks}
We gratefully acknowledge support from the New York University Tandon Center for Cybersecurity, the Max Planck Institute for Software Systems, and the University of Maryland.
The third author conducted this work while at the Max Planck Institute for Software Systems.
\end{acks}
\bibliographystyle{ACM-Reference-Format}
\bibliography{ms}
\appendix

\section{Meme}
\label{appendix:meme}

\begin{figure}[ht]
    \centering
    \includegraphics[width=0.4\linewidth]{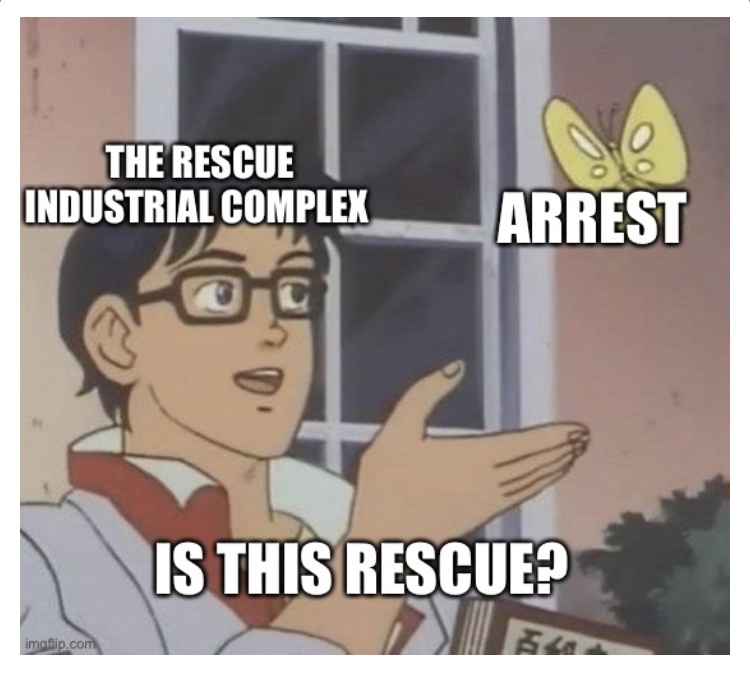}
    \caption{A sex worker meme about the rescue industry's misunderstanding of sex work. It shows that while some members of the rescue industry may believe that arrest can help ``rescue'' sex workers, sex workers know it to not be the case.}
    \label{fig:meme}
\end{figure}

% \begin{figure}[ht]
%     \centering
%     \begin{minipage}{.47\textwidth}
%         \centering
%         \includegraphics[width=0.95\linewidth]{images/meme1.jpeg}
%     \end{minipage}%
%     \hfill
%     \begin{minipage}{.47\textwidth}
%         \centering
%         \includegraphics[width=0.95\linewidth]{images/meme2.jpeg}
%     \end{minipage}
%     \caption{Two sex worker memes about the rescue industry.}
%     \label{fig:memes}
% \end{figure}

\section{Organization Representative Interview Protocol}
\label{org_rep_protocol}

\noindent Intro: [Review consent and confirm recording]

\noindent Background / Warm up

\begin{itemize}
    \item Tell me a little about your background? What type of social work do you do?
    \item How did you start working with this organization? 
    \item Have you been working with \sas? If so, how long?
\end{itemize}

\noindent Impact of the Organization

\begin{itemize}
    \item How many full time workers and how many volunteers do you have in your organization?
    \item Whom do you serve? (How many are being trafficked?) (Where do you find them?)
    \item How many people do you serve?
    \item Does it change with adoption of online outreach?
\end{itemize}

\noindent Experience with Communication

\begin{itemize}
    \item What are your goals when reaching out to people whom you suspect might be victims of human trafficking? 
    \begin{itemize}
        \item How do your goals affect the method of outreach?
        \item How do they affect technology use?
        \item How do they affect your approach to identity/circumstances?
        \item Do you use a ``walking with'' period to determine if you can give help or not?
    \end{itemize}
    \item What kinds of things do you think about when you attempt to reach people? 
    \item Can you describe the types of people and/or scenarios you encounter?
    \begin{itemize}
        \item How much of the time (on average) is the recipient a trafficker? 
        \item How much of the time (on average) do you encounter spam?
        \item How do you decide whether to continue pursuing the number, based on the initial response?
        \item Have you ever encountered minors? Do you respond differently if the person is a minor?
    \end{itemize}
    \item When you do in fact reach a trafficking victim, what problems have you encountered?
    \item Typically, when you reach out, where are you located? How does your location impact your work?
\end{itemize}

\noindent Knowledge of Sex \add{Industry}

\begin{itemize}
    \item What, if any, role does the identity of those you help play in your outreach?
    \item What about in terms of the types of problems you encounter?
    \item Does it change the type of help you offer, or are able to offer?
\end{itemize}

\noindent Perception of Online Outreach

\begin{itemize}
    \item (If they are using \sas\ platform) How long have you been using the platform?
    \begin{itemize}
        \item What has been your experience with it?
        \item What do you like about it?
        \item What do you dislike?
    \end{itemize}
    \item How does using this platform impact how you reach out?
    \begin{itemize}
        \item What benefits does it have? For whom / what circumstances?
        \item What challenges / risks does it have? For whom / what circumstances?
    \end{itemize}
    \item \sas's platform uses SMS text messaging to reach out. How do you feel about using SMS text messaging?
    \begin{itemize}
        \item Do you use other types of messaging apps outside of your work with \sas?
        \item Are there safety concerns with this that you don’t encounter with other messaging apps, or vice versa? What about with social media or printed messages?
    \end{itemize}
\end{itemize}

\noindent Other Outreach Methods

\begin{itemize}
    \item How / when are you using technologies (video conferencing, mobile, etc) alongside \sas's, and for whom?
    \begin{itemize}
        \item Do you give them your personal number (so they trust you / can call you back when they are safe) or would you say it is unsafe to give out your number?
    \end{itemize}
    \item What resources other than the platform do you use to identify victims?
    \begin{itemize}
        \item Do you have any networks that you work with?
        \item Who is involved in this network?
        \item How do you keep in touch / communicate with that network?
    \end{itemize}
    \item What other methods besides this system do you currently use to reach out to potential victims? Probe for each method:
    \begin{itemize}
        \item What are the circumstances in which you would use that method? 
        \item What are the benefits? 
        \item What are the disadvantages? 
        \item How does [this method] compare with \sas’s platform?
    \end{itemize}
    \item How have you adapted your outreach process to take it online?
    \begin{itemize}
        \item Are there any other methods that you have used in the past that you don’t now?
    \end{itemize}
    \item How would you compare the effectiveness of online outreach vs. street outreach?
\end{itemize}

\noindent Security and Privacy

\begin{itemize}
    \item Has reaching out ever affected the safety of the person whom you are contacting? If so, how? Could you give me an example?
    \begin{itemize}
        \item How did the person respond? 
        \item Has this changed the way you reach out? 
        \item Has this experience changed your overall approach?
        \item Has your approach changed over time? 
        \item If so, how has it changed? 
        \begin{itemize}
            \item Is there anything that you used to do that you no longer do?
            \item Is there anything that you considered doing but decided not to? Why not?
        \end{itemize}
        \item Can you think of other examples that are relevant? 
    \end{itemize}
    \item Has reaching out ever affected your own safety? Explain.
    \item What efforts or steps do you take to keep yourself safe? 
    \begin{itemize}
        \item How do you avoid giving away your location? Is that important? Why / why not?
        \item What other privacy / security measures do you take?
        \item How successful do you think you are? 
        \item Do you still have concerns? Explain.
    \end{itemize}
    \item Do you have any safety concerns with the platform? Probe for each: 
    \begin{itemize}
        \item How do you address that concern?
        \item How successful do you think you are? 
    \end{itemize}
    \item Do you still have concerns? Explain.
    \item How do you handle your work with respect to victims’ traffickers?
    \begin{itemize}
        \item How do you attempt to keep your communications secret from potential victims' traffickers when you reach out? Are you always successful? Why / why not? How do you know?
        \item Could you describe an instance when you were not successful? What happened? 
        \item How did that influence your methods for reaching out?
        \item Has this experience changed your approach?
        \item Has your approach changed over time? 
        \item If so, how has it changed? 
        \begin{itemize}
            \item Is there anything that you used to do that you no longer do?
            \item Is there anything that you considered doing but decided not to? Why not?
        \end{itemize}
        \item Can you think of other examples that are relevant? 
    \end{itemize}
\end{itemize}

\noindent Is there anything else you would like me to know?
\section{\sas\ Employee Interview Protocol}
\label{sas_protocol}

\noindent Intro: [Review consent and confirm recording]

\noindent Background / Warm up

\begin{itemize}
    \item Tell me about your background?
    \item What is your role at \sas?
    \begin{itemize}
        \item How long have you been in this role?
        \item Are you a volunteer or paid?
        \item What software have you worked on?
    \end{itemize}
\end{itemize}

\noindent Tell me about \sas's mission and goals?

\noindent How might someone use \fs?

\noindent What types of features / tools does \fs provide?

\begin{itemize}
    \item Tell me about these tools?
    \item What are they for?
    \item How would someone use it? Could you give me an example?
\end{itemize}

\noindent What demographic features / tools does \fs provide?

\begin{itemize}
    \item Specifically, what types of information?
    \item Why did you design it that way?
    \item Can people use the platform to identify people based on their demographic information? How would they do that?
\end{itemize}

\noindent What safety features / tools does \fs provide? Why?

\noindent What considerations went into the design of \fs?

\begin{itemize}
    \item Probe: safety, demographics, statistics / reports, other?
    \item (Contrast with Project Intercept if you can)
\end{itemize}

\noindent What needs is \fs designed to address?

\begin{itemize}
    \item Probe: safety, demographics, statistics / reports, other?
    \item (Make sure you understand when a feature / tool is entirely new vs upgrade, etc)
\end{itemize}

\noindent What are some challenges that organizations that use your platform have?

\begin{itemize}
    \item How did you learn about that challenge? 
    \item What were most critical to address? Why?
    \item How did you address those challenges?
    \item Which do you think are easiest/hardest to address?
    \item Probe: What about safety? 
\end{itemize}

\noindent What types of volunteer organizations does \sas\ support?

\begin{itemize}
    \item What are some typical types of problems they are trying to address?
\end{itemize}

\noindent What types of products does \sas\ provide to these organizations?

\begin{itemize}
    \item Why did you design these products?
    \item What are they used for?
    \item For each: What types of organizations (that you have just described) use these products? How?
\end{itemize}

\noindent How do you learn about the needs and requirements of organizations that use your product?

\begin{itemize}
    \item How do they communicate needs to \sas? 
    \item Probe: Do you conduct research? Meetings? Analytics? What else? Could you tell me more about what role these forms of inquiry play?
    \item How do you communicate those ideas to developers? Are developers involved in the process? 
    \item Are there product teams? 
\end{itemize}
\section{Sex Worker Interview Protocol}
\label{sw_protocol}

\noindent Intro: 
\begin{itemize}
    \item The goal of our interviews is to understand the ways support is currently given, and improve the ways it could be given.
    \item (Review consent and confirm recording)
    \item Emphasize that you do not need to answer any question or describe any negative experience you don't want to and will still get the gift card)
\end{itemize}

\noindent (Tests to see if they don't have experience with sex work in North America)

\begin{itemize}
    \item If you determine they are a scammer, ask 2 or 3 more questions before ending the interview with ``Thank you.'' This way the scammers don’t learn our tests.
\end{itemize}

\noindent Background

\begin{itemize}
    \item Tell me a little about your background? What type of work do you do?
    \item How did you get into sex work?
    \item If you're comfortable sharing, how long have you been doing sex work?
\end{itemize}

\noindent Communication

\begin{itemize}
    \item How do / did you find clients?
    \item (If appropriate) What methods do / did you use for initial and continuing communication with clients?
    \item (If appropriate) What is / was the process for scheduling an appointment? Do / did you require a deposit? 
    \item (If appropriate) Is / was there somebody who coordinates some things for you, or do / did you do everything yourself?
    \item (If appropriate) Where are / were appointments held?
\end{itemize}

\noindent Experience with Reachouts

\begin{itemize}
    \item Have you ever been approached by an organization or person who wanted to offer help or services? (e.g., medical, loan, housing, immigration, other services)
    \item If so: Could you tell me about that? 
    \begin{itemize}
        \item How were you contacted?
        \item Who contacted you?
        \item What was your reaction? 
        \item If not positive reaction: What would you have preferred to have happen?
        \item At any time, did you switch methods of communication (e.g., text to phone)? Why?
        \item Did you end up using the resources offered? Why/why not?
        \begin{itemize}
            \item What helped you? 
            \item How long did it take to get from initial contact to receiving help?
            \item If it took a long time: Was the period of waiting initiated by you or your contact? Was it helpful?
        \end{itemize}
        \item How do you think this service found you? 
        \begin{itemize}
            \item How do you feel about them finding you this way?
            \item Do you think they used information from your ads to contact you? How do you feel about them using that information?
        \end{itemize}
        \item Are there times you can think of / have seen where it would be unsafe for you if an organization reached out in-person / text / online?
        \begin{itemize}
            \item What can the person reaching out do to mitigate this? What do you think is the best way for them to reach out?
            \item What do you think is the safest way to make contact? Why?
            \item What do you think is the best way to make contact? Why?
        \end{itemize}
        \item Are there things that those doing outreach should avoid saying or doing when reaching out?
        \begin{itemize}
            \item In the content of text messages? What about in-person?
        \end{itemize}
        \item What made you trust that the person/org that reached out to you was real?
        \begin{itemize}
            \item Were you worried that they were police?
        \end{itemize}
        \item Are there some people/organizations you are more comfortable reaching out to you vs. others (e.g., religious, non religious)?
        \item Re-prompt about any other similar experiences or experiences of others they have heard about
    \end{itemize}
\end{itemize}

\noindent Experience with seeking services / support

\begin{itemize}
    \item Have there been instances where you needed support or services? (e.g., medical, loan, housing, immigration, other services)
    \item If you're comfortable sharing, what types of services did you need?
    \item How did you go about finding those services? 
    \item Do you know of any organizations providing the kind of support that you needed, that you didn't reach out to, and if yes, why did you choose to not approach them? 
    \item Are you aware of other organizations providing other kinds of help? 
    \item Do you think there is something the organizations providing support could have done to make it easier for you or others to find / access the services you needed?
    \item Have you ever gone to someone other than these organizations for help? For example, a client, a colleague, family?
    \item If it has not come up yet:
    \begin{itemize}
        \item Imagine that an organization that provides services wanted to reach out to you proactively. How would you feel about that? 
        \item What would you want them to say or offer?
        \item How would you expect them to find you?
        \item Are there times you can think of / have seen where it would be unsafe for you if an organization reached out in-person / text / online?
        \begin{itemize}
            \item What can the person reaching out do to mitigate this? What do you think is the best way for them to reach out?
            \item What do you think is the safest way to make contact? Why?
            \item What do you think is the best way to make contact? Why?
        \end{itemize}
        \item Are there things that those doing outreach should avoid saying or doing when reaching out?
        \begin{itemize}
            \item In the content of text messages? What about in-person?
        \end{itemize}
        \item What would make you trust that the person / org that reached out to you was real?
        \begin{itemize}
            \item Would you be worried that they were police?
        \end{itemize}
        \item Are there some people / organizations you are more comfortable reaching out to you vs. others (e.g., religious, non religious)?
    \end{itemize}
\end{itemize}

\noindent Experience with doing outreach

\begin{itemize}
    \item Have you ever reached out to another sex worker to offer help or assistance?
    \begin{itemize}
        \item How did you know or identify that your assistance might be useful to them?
        \item How was your help received? Do you think being a peer made a difference to that?
        \item Are you aware of any organizations within the sex worker community that offer help to sex workers? Can you tell me a little bit about those organizations?
        \begin{itemize}
            \item How do you feel like these organizations differ from external organizations offering services to sex workers?
            \item In what ways do you feel like these organizations are similar to external organizations offering services to sex workers?
        \end{itemize}
    \end{itemize}
\end{itemize}

\noindent Identity

\begin{itemize}
    \item Do you think that your identity (e.g., your sexual identity, gender, race, or some other characteristic) is important to the types of support you receive?
    \item What about the identity of who reaches out to you and what they say?
    \item If applicable: do you remember if during any of the outreach you received, your identity was discussed? Would that have been meaningful to you, how? What about the identity of the person reaching out?
    \item Do you think it's important when receiving help that the person or organization reaching out knows (or does not know) information about your identity? What kinds of things? Why?
    \begin{itemize}
        \item Probe: gender, sexual identity, type of work, age, race, location, other
    \end{itemize}
\end{itemize}

\noindent Is there anything else you would like me to know about the process of seeking support or resources, or having people reach out to you offering those resources?

\end{document}